\documentclass[sigconf, nonacm]{acmart}

\usepackage{hyperref}
\usepackage[ruled,vlined]{algorithm2e}
\usepackage{amsmath}

\newtheorem{problem}{Problem}

\begin{document}

\title{Three-dimensional Geospatial Interlinking with JedAI-spatial}

\author{Marios Papamichalopoulos$^1$, George Papadakis$^1$,  George Mandilaras$^1$, Maria Despoina Siampou$^1$, Nikos Mamoulis$^2$, Manolis Koubarakis$^1$}
\affiliation{%
  \institution{$^1$National \& Kapodistrian University of Athens, Greece $\>\>$ {\{cs3190006, gpapadis, gmandi, m.siampou, koubarak\}@di.uoa.gr}\\
  $^2$University of Ioannina, Greece $\>\>$ nikos@cs.uoi.gr\\
  \country{}
  }
}

\renewcommand{\shortauthors}{Papamichalopoulos, Papadakis, Mandilaras, Siampou, Mamoulis, Koubarakis}

\begin{abstract}
Geospatial data constitutes a considerable part of (Semantic) Web data, but so far, its sources are inadequately interlinked in the Linked Open Data cloud. Geospatial Interlinking aims to cover this gap by associating geometries with topological relations like those of the Dimensionally Extended 9-Intersection Model. Due to its quadratic time complexity, various algorithms aim to carry out Geospatial Interlinking efficiently. We present \textit{JedAI-spatial}, a novel, open-source system that organizes these algorithms according to three dimensions:  (i) \textit{Space Tiling}, which determines the approach that reduces the search space, (ii) \textit{Budget-awareness}, which~distinguishes interlinking algorithms into batch and progressive ones, and (iii) \textit{Execution mode}, which discerns between serial algorithms, running on a single CPU-core, and parallel ones, running on top of Apache Spark. We analytically describe JedAI-spatial's architecture and capabilities and perform thorough experiments to provide interesting insights about the relative performance of its algorithms.
\end{abstract}




\maketitle

\section{Introduction}

Geospatial data has escalated tremendously over the years. The outbreak of Internet of Things (IoT) devices, smartphones, position tracking applications and location-based services has skyrocketed the volume of geospatial data. For example, 100TB of weather-related data is produced everyday\footnote{\url{https://www.ibm.com/topics/geospatial-data}}; Uber hit the milestone of 5 billion rides among 76 countries already on May 20, 2017\footnote{\url{https://www.uber.com/en-SG/blog/uber-hits-5-billion-rides-milestone}}. Web platforms like OpenStreetMap\footnote{https://www.openstreetmap.org} provide an open and editable map of the whole world. Earth observation programmes like Copernicus\footnote{\url{https://www.copernicus.eu}} publish tens of terabytes of geospatial data per day on the Web\footnote{{\scriptsize\url{https://www.copernicus.eu/sites/default/files/Copernicus_DIAS_Factsheet_June2018.pdf}}}. For these reasons, geospatial data constitutes a considerable part of Semantic Web data, but the links between its data sources and their geometries are scarce in the Linked Open Data cloud \cite{DBLP:conf/semweb/Ngomo13,DBLP:conf/www/0001MMK21}.

\begin{figure}[t]
\centering
\includegraphics[width=0.47\textwidth]{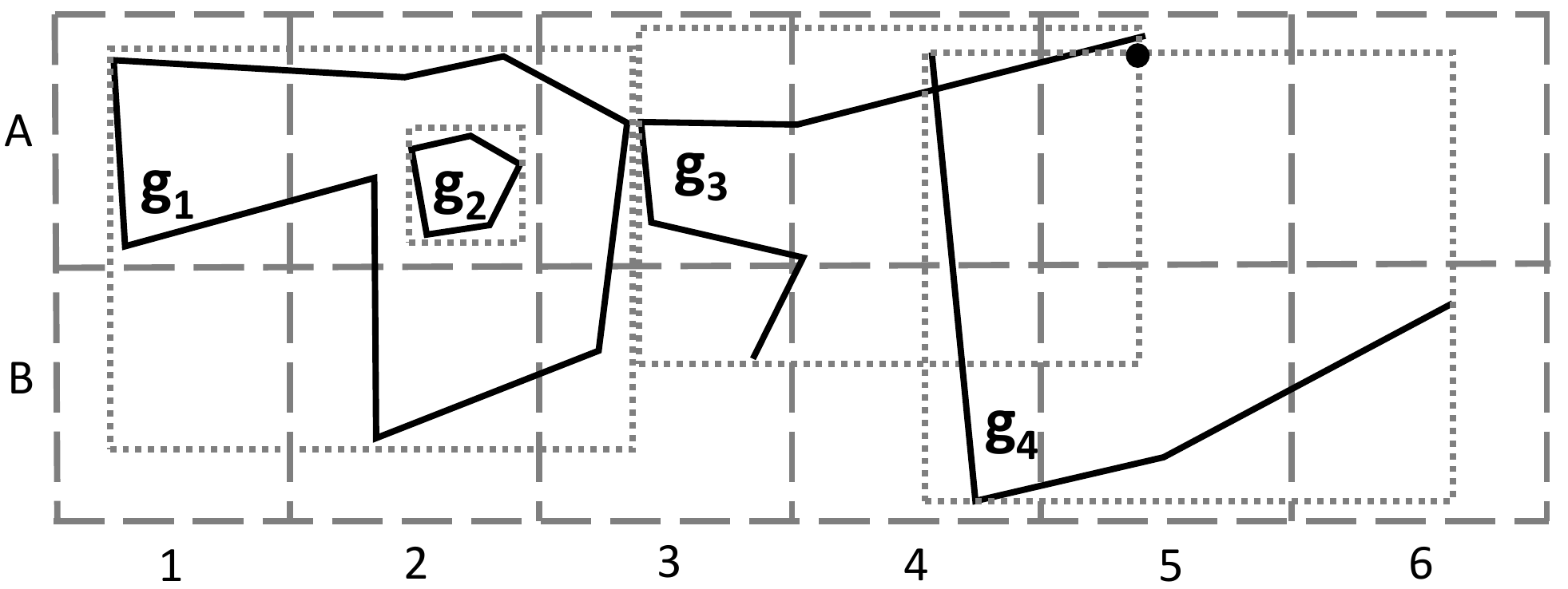}
\vspace{-5pt}
\caption{Example of four topologically related geometries. 
}
\vspace{-18pt}
\label{fig:example}
\end{figure}

\textit{Geospatial Interlinking} aims to cover this gap by associating pairs of geometries with topological relations like those of the Dimensionally Extended 9-Intersection Model (DE-9IM) \cite{egenhofer1991point,DBLP:conf/ssd/ClementiniFO93,DBLP:journals/cg/ClementiniSE94}. In Figure \ref{fig:example} for instance, LineString $g_3$ intersects LineString $g_4$ and touches Polygon $g_1$, which contains Polygon $g_2$. Two are the main challenges of this task: (i) its inherently quadratic time complexity, because it has to examine every pair of geometries, and
(ii) the high time complexity of examining a single pair of geometries, which amounts to $O(N \log N)$, where $N$ is the size of the union set of their boundary points \cite{DBLP:conf/ssd/ChanN97}. As a result, Geospatial Interlinking involves a high computational cost that does not scale to large Web datasets.

Numerous algorithms aim to address these challenges by enhancing the time efficiency and scalability of Geospatial Interlinking. 
The most recent ones operate in main memory, reducing the search space to pairs of geometries that are likely to be topologically related according to a geospatial index \cite{DBLP:journals/pvldb/SowellSCDG13,DBLP:journals/pvldb/SidlauskasJ14,DBLP:journals/pvldb/PandeyKNK18}. However, no open-source system organizes these algorithms into a common framework so as to facilitate researchers and practitioners in their effort to populate the LOD cloud with more topological relations. Systems like Silk \cite{DBLP:conf/semweb/JentzschIB10} and LIMES \cite{DBLP:conf/ijcai/NgomoA11} convey only the methods developed by their creators, Silk-spatial \cite{DBLP:conf/www/SmerosK16} and RADON \cite{DBLP:conf/aaai/SherifDSN17} respectively, while systems that could act as a library of established methods, such as stLD \cite{DBLP:conf/sigmod/SantipantakisGD19,DBLP:books/sp/20/SantipantakisDVV20}, are not publicly available. Moreover, no system supports progressive methods, neither for serial nor for parallel processing, even though they are indispensable for applications with limited computational or temporal resources \cite{DBLP:conf/www/0001MMK21}.


To address these issues, we present \textit{JedAI-spatial}, an open-source system that supports a broad range of Geospatial Interlinking applications on the Web by implementing the state-of-the-art methods in the literature. JedAI-spatial makes the following contributions:

\noindent
$\bullet$ It organizes the main algorithms into a novel taxonomy that facilitates their use and adoption by practitioners and researchers, enabling them to select the most appropriate algorithm based on the requirements of their application.

\noindent
$\bullet$ Its intuitive user interface supports both novice and expert users.

\noindent
$\bullet$ Its modular and extensible architecture allows for easily incorporating new algorithms and improvements to the existing ones.

\noindent
$\bullet$ It optimizes the implementation of existing algorithms, some of which have not been applied to Geospatial Interlinking before.


\noindent
$\bullet$ We have publicly released the code of JedAI-spatial at:\\ \url{https://github.com/giantInterlinking/JedAI-spatial}.


The rest of the paper is structured as follows: Section \ref{sec:relatedWork}  refers to the main works in the literature, Section \ref{sec:preliminaries} provides background knowledge on Geospatial Interlinking, while Section \ref{sec:architecture} describes JedAI-spatial's architecture, 
explaining the role of every component. 
Section \ref{sec:backEnd} delves into its back-end, outlining the functionality of every supported method and highlighting our improvements that lead to significantly higher time efficiency. We briefly describe its front-end in Section \ref{sec:gui}, we discuss the Web applications that could benefit from JedAI-spatial in Section \ref{sec:applications}, we perform an experimental analysis over large, real datasets in Section \ref{sec:quantAnalysis},
and provide a qualitative analysis of the main supported methods in Section \ref{sec:qualitativeAnalysis}. 

\section{Related Work}
\label{sec:relatedWork}

In the Semantic Web domain, there are three related systems:
\begin{enumerate}
    \item \textit{Silk} \cite{DBLP:conf/semweb/JentzschIB10} constitutes an open-source, generic framework for Link Discovery that comprises a specialized component for Geospatial Interlinking, called Silk-spatial \cite{DBLP:conf/www/SmerosK16}. It exclusively supports a budget-agnostic, parallel method that runs on top of Apache Hadoop\footnote{\url{http://hadoop.apache.org}}. Its Filtering relies on a static, coarse-grained Equigrid, whose dimensions are defined by the user. Its Verification computes one topological~relation per~run.
    \item \textit{LIMES} \cite{DBLP:conf/ijcai/NgomoA11} is an open-source, generic framework for Link Discovery with two algorithms for Geospatial Interlinking: ORCHID \cite{DBLP:conf/semweb/Ngomo13}, which detects proximity relations in an efficient way, and RADON \cite{DBLP:conf/aaai/SherifDSN17}, which detects topological relations. RADON's Filtering employs a dynamic Equigrid, whose granularity depends on the input data, while~its~Verification employs a hash map that maintains all examined pairs in memory to avoid repeated computations. Due to this data structure, RADON has been parallelized as a multi-core, shared-memory process, rather than a shared-nothing, MapReduce-based approach.
    Originally, its Verification computed a single topological relation per run, but was later extended to compute all relations at once \cite{DBLP:conf/semweb/AhmedSN18}.
    \item \textit{stLD} \cite{DBLP:conf/sigmod/SantipantakisGD19,DBLP:books/sp/20/SantipantakisDVV20} is a tool that is crafted for Geospatial Interlinking. It is limited to budget-agnostic approaches, conveying a variety of algorithms, such as R-Tree, static Equigrid as well as hierarchical grid. Similar to JedAI-spatial and GIA.nt \cite{DBLP:conf/www/0001MMK21}, its algorithms are capable of loading only the source dataset in main memory, reading the target one on-the-fly. stLD also supports massive parallelization on top of Apache Flink\footnote{\url{http://flink.apache.org}}. Its Verification supports both proximity and topological relations, but computes a single relation per run. However, its code has not been publicly released.
\end{enumerate}

Note that none of these systems supports budget-aware algorithms, unlike JedAI-spatial, which conveys the existing progressive methods \cite{DBLP:conf/www/0001MMK21,DBLP:conf/tsas/0001MMK22} along with new ones. 
Note also that these systems are of limited scope, as the open-source systems (Silk and LIMES) convey only the algorithms developed by their creators, while stLD is a proprietary software that cannot be used as a library of state-of-the-art tools in Web applications nor can be extended with novel techniques and pipelines.

To the best of our knowledge, no other systems similar to JedAI-spatial have been publicly released. The most relevant tools, which support parallelization on top of Apache Hadoop or Spark, are analyzed in \cite{DBLP:journals/pvldb/PandeyKNK18}. They all support a variety of spatial queries, such as distance (range) and kNN queries, but in the context of Geospatial Interlinking, only their spatial join is applicable. Each tool essentially offers a single parallel algorithm for this join. The most recent and advanced systems are GeoSpark \cite{DBLP:conf/gis/YuWS15} (a.k.a., Apache Sedona), Spatial Spark \cite{DBLP:conf/icde/YouZG15}, Location Spark \cite{DBLP:journals/pvldb/TangYMOA16,DBLP:journals/fdata/TangYMMOA20} and Magellan. All these algorithms have been integrated into JedAI-spatial.

Finally, related to JedAI-spatial are two works that examine the relative performance of 10 serial, budget-agnostic algorithms that run in main memory
\cite{DBLP:journals/pvldb/SowellSCDG13,DBLP:journals/pvldb/SidlauskasJ14}. However, their experimental analyses focus on answering distance (range) queries about moving objects.
\section{Preliminaries}
\label{sec:preliminaries}

JedAI-spatial supports two types of geometries:
\begin{enumerate}
    \item the one-dimensional \textit{LineStrings} or \textit{Polylines}, which comprise a sequence of points and the line segments that connect the consecutive ones (e.g., $g_3$ and $g_4$ in Figure \ref{fig:example}),
    \item the two-dimensional \textit{Polygons}, which usually comprise a sequence of connected points, where the first and last one coincide (e.g., $g_1$ and $g_2$ in Figure \ref{fig:example}).
\end{enumerate}

Both types of geometries consist of an interior, a boundary and an exterior (i.e., all points that are not part of the interior or the boundary). These three parts are used by the DE9IM model, which has been standardized by the Open Geospatial Consortium (OGC), to define 10 topological relations between two geometries $A$ and $B$:

\begin{enumerate}
    \item \texttt{equals}$(A, B)$: the interiors and boundaries of $A$ and $B$ are identical.
    \item \texttt{disjoint}$(A, B)$: $A$ and $B$ have no point in common, as the interior and boundary of $A$ intersect neither with the interior nor with the boundary of $B$. 
    \item \texttt{intersects}$(A, B)$: $A$ and $B$ have at least one point in common, i.e., their interiors or boundaries are not \texttt{disjoint}.
    \item \texttt{touches}$(A, B)$: the boundaries of $A$ and $B$ intersect but their interiors do not.
    \item \texttt{within}$(A, B)$: $A$ is located inside the interior of $B$.
    \item \texttt{contains}$(A, B)$: \texttt{within}$(B, A)$.
    \item \texttt{covers}$(A, B)$: all points of $B$ lie in $A$'s interior or boundary.
    \item \texttt{covered-by}$(A, B)$: \texttt{covers}$(B, A)$.
    \item \texttt{crosses}$(A, B)$: $A$ and $B$ have some interior points in common but not all, while $dim(A)$$<$$dim(B)$ or $dim(B)$$<$$dim(A)$.
    \item \texttt{overlaps}$(A, B)$: $A$ and $B$ have some points in common but not all, while $dim(A)=dim(B)$.
\end{enumerate}

Note that $dim(g)$ amounts to 0, 1 or 2 if geometry $g$ is a point, a line segment or an area, respectively. Note also that JedAI-spatial disregards the relation \texttt{disjoint} for two reasons \cite{DBLP:conf/www/0001MMK21}: (i) it provides no positive information for the relative location of two geometries, and (ii) it is impractical to compute it in the case of large input data, because it scales quadratically with the input size, given that the vast majority of geometries are disjoint. Yet, JedAI-spatial relies on a closed-world assumption: the lack of the relation \texttt{intersects} between two geometries implies that they satisfy the relation \texttt{disjoint}.

Following \cite{DBLP:conf/www/0001MMK21,DBLP:conf/semweb/AhmedSN18}, JedAI-spatial considers \textit{Holistic Geospatial Interlinking}, which simultaneously computes all \textbf{positive} topological relations (i.e., all DE9IM relations except for \texttt{disjoint}): for each pair of geometries, it estimates the Intersection Matrix, from which all relations can be extracted with simple boolean expressions\footnote{See \url{https://en.wikipedia.org/wiki/DE-9IM\#Matrix_model} for more details.}. In contrast, Silk, LIMES and stLD compute an individual topological relation per run, 
repeating the entire processing to the same data multiple times in order to produce all topological links.

Overall, Geospatial Interlinking is formally defined as:

\begin{problem}[Geospatial Interlinking]
Given a source and a target dataset, $S$ and $T$, together with the set of positive topological relations $R$, compute the set of links $L_R = \{ (s, r, t) \subseteq S \times R \times T : r(s, t) \}$ from the Intersection Matrix of all related geometry pairs.
\label{pr:batchGI}
\end{problem}

Given that all batch Geospatial Interlinking algorithms produce an \underline{exact} solution, yielding the same links, their performance is exclusively assessed with respect to time efficiency, i.e., run-time.

\begin{figure}[t]
\centering
\includegraphics[width=0.35\textwidth]{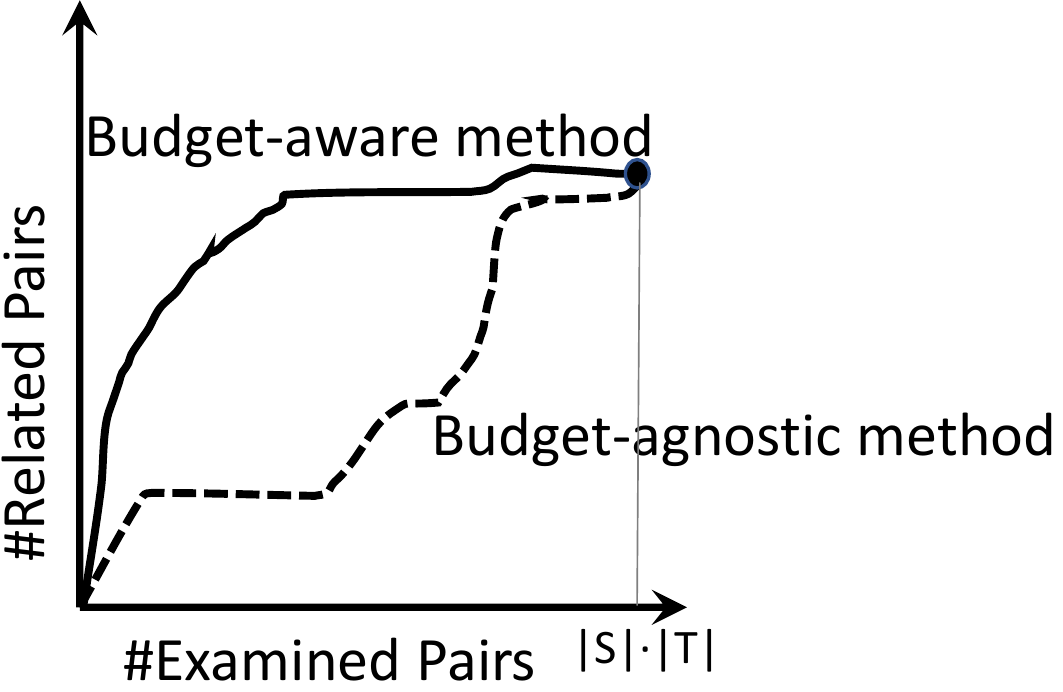}
\vspace{-5pt}
\caption{The Progressive Geometry Recall of budget-agnostic (batch) and budget-aware (progressive) algorithms.}
\vspace{-10pt}
\label{fig:pgrExample}
\end{figure}

\textbf{Progressive Geospatial Interlinking.} An \underline{approximate} solution to Geospatial Interlinking is provided by progressive algorithms, which run for a limited time or number of calculations. These algorithms are necessary for applications with limited resources, such as
cloud-based applications with a specific budget for AWS Lambda functions, which charge whenever they are called~\cite{DBLP:journals/soca/VillamizarGOCSV17}. 

Compared to batch algorithms, the goal of progressive algorithms is twofold  \cite{DBLP:conf/www/0001MMK21}: (i) they should produce the same results if they process the entire input data, and (ii) they should detect a significantly larger number of related geometry pairs, if their operation is terminated earlier. 

These requirements are reflected in Figure \ref{fig:pgrExample}, where the horizontal axis corresponds to the number of examined pairs and the vertical one to the number of related pairs. Essentially, the progressive algorithms should define a processing order that examines the related pairs before the non-related ones, unlike batch algorithms, which examine pairs in an arbitrary order. Hence, the progressive algorithms should maximize the area under their curve, an evaluation measure called \textbf{Progressive Geometry Recall} that is defined in $[0, 1]$, with higher values indicating higher effectiveness.

Given a budget $BU$ on the maximum calculations or running time, progressive algorithms tackle the following task~\cite{DBLP:conf/www/0001MMK21}:
\begin{problem}[Progressive Geospatial Interlinking]
Given a source and a target dataset, $S$ and $T$, the positive topological relations $R$ and a budget $BU$, maximize Progressive Geometry Recall within~$BU$.
\label{pr:progressiveGI}
\end{problem}
\vspace{-10pt}
Progressive algorithms are also evaluated with respect to: (i) run-time, (ii) \textit{precision}, i.e., the number of detected related pairs divided by the number of examined pairs, and (iii) \textit{recall}, i.e., the number of detected related pairs divided by the maximum number of related pairs that would be found within $BU$ examinations in the optimal case, after placing all related pairs before the non-related~ones.
\section{System Architecture}
\label{sec:architecture}

JedAI-spatial organizes the Geospatial Interlinking algorithms into a novel taxonomy formed by three dimensions, as shown in Figure~\ref{fig:spaceTiling}:
\begin{enumerate}
    \item \textit{Space Tiling} distinguishes the Geospatial Interlinking algorithms into \textit{grid-}, \textit{tree-} and \textit{partition-based} ones. The first category includes Semantic Web techniques that define a static or dynamic Equigrid, the second one encompasses main-memory spatial join techniques from the database community, and the third one conveys variations of plane sweep, a cornerstone algorithm of computational geometry.
    \item \textit{Budget-awareness} categorizes algorithms into \textit{budget-agnostic} and \textit{budget-aware} ones. The former are executed in a batch manner that processes the input data in no particular order and produces results only upon completion of the entire process. Budget-aware algorithms are suitable for applications with limited computational or temporal resources, producing results progressively, in a pay-as-you-go manner.
    \item \textit{Execution mode} distinguishes between serialized algorithms, which run on a single CPU core, and massively parallel ones, which run on top of Apache Spark {\small (\url{https://spark.apache.org})}.
\end{enumerate}

\begin{figure}[t]
\centering
\includegraphics[width=0.47\textwidth]{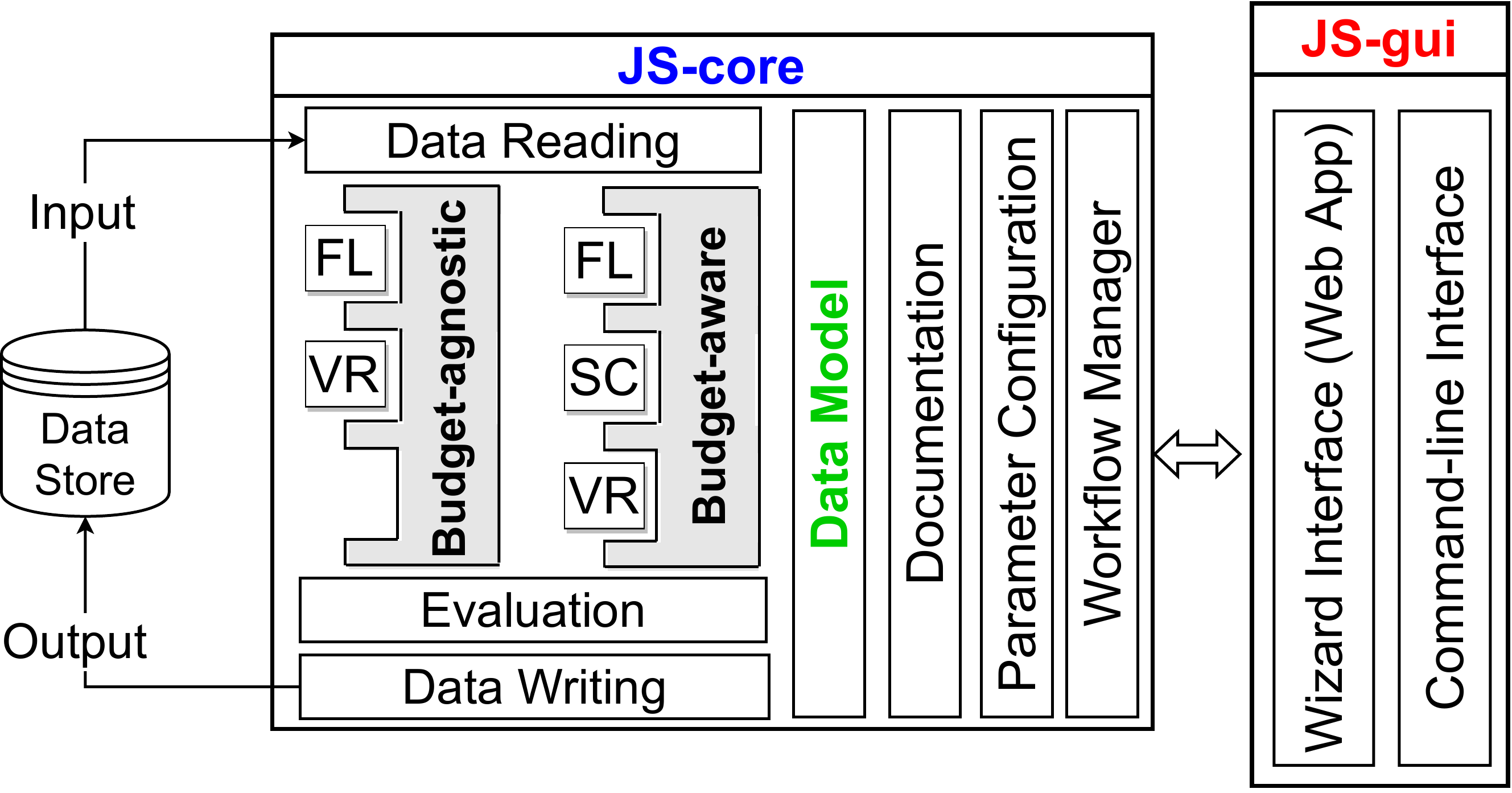}
\vspace{-10pt}
\caption{{\small The model-view-controller architecture of JedAI-spatial.}}
\vspace{-15pt}
\label{fig:architecture}
\end{figure}

JedAI-spatial creates any end-to-end pipeline that is defined by these three dimensions. This is achieved by the model-view-controller architecture in Figure \ref{fig:architecture}:
JS-gui offers two interfaces for user interaction (view), JS-core conveys numerous algorithms and pipelines (controller), and the Data Model component provides the data
structures that lie at its core (model). 

This architecture serves the following goals:


\begin{figure}[t]
\centering
\includegraphics[width=0.44\textwidth]{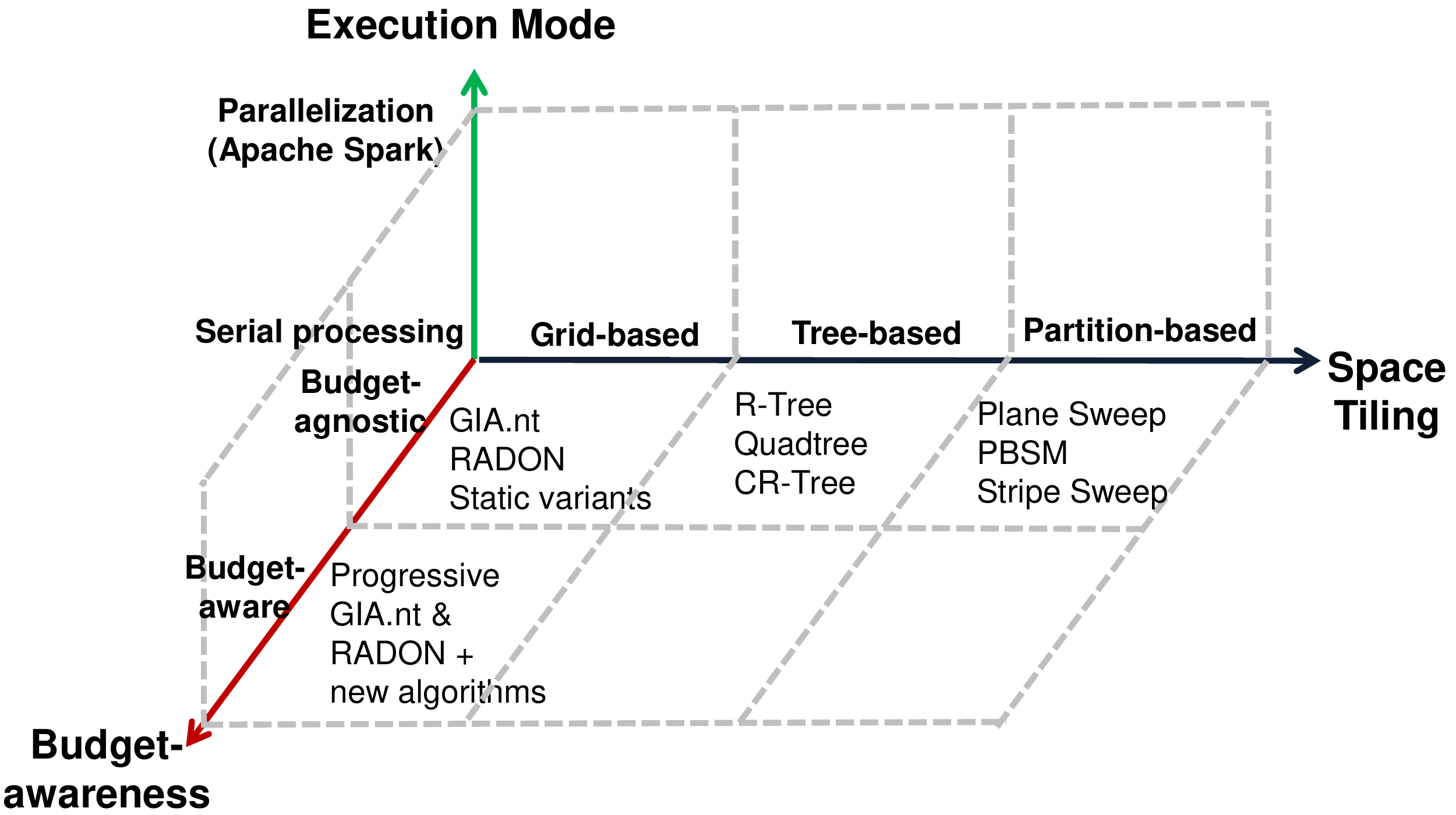}
\vspace{-8pt}
\caption{The solution space of Geospatial Interlinking algorithms that can be constructed by JedAI-spatial.}
\vspace{-18pt}
\label{fig:spaceTiling}
\end{figure}


\vspace{3pt}
\noindent
$\bullet$ \textit{Broad data coverage.} Through its \textit{Data Reading} component, JedAI-spatial supports the most popular structured and semi-structured formats that are used for encoding geometries: Well Known Text (WKT) files, CSV and TSV files, GeoJSON files, RDF dumps,  JsonRDF as well as SPARQL endpoints. In this way, JedAI-spatial is able to interlink heterogeneous datasets, e.g., WKT data with GeoJSON files. Special care has been taken to detect and ignore corrupted data as well as to remove noise, which is quite common in automatically or user-generated data sets, especially as their size grows.

\vspace{3pt}
\noindent
$\bullet$ \textit{Broad algorithmic coverage.} JedAI-spatial serves as a library of the state-of-the-art algorithms in the literature, even if they haven't been applied directly to Geospatial Interlinking before. These algorithms are implemented by JS-core (cf. Section~\ref{sec:backEnd}).

\vspace{3pt}
\noindent
$\bullet$ \textit{Broad application coverage.} JedAI-spatial accommodates both academic and commercial applications, as its code is released under Apache License V2.0. It also supports both batch and progressive applications. In any type of applications, it is crucial to detect the most suitable algorithm for the data at hand (e.g., different batch algorithms might be faster in a LineString-to-LineString scenario than in a Polygon-to-LineString one). To cover this need, JedAI-spatial's benchmarking functionality facilitates the comparative evaluation 
of a large variety of pipelines.

\vspace{3pt}
\noindent
$\bullet$ \textit{High usability.} JedAI-spatial supports both novice and expert users. The former can apply complex, high performing pipelines to their data simply by choosing among the available algorithms, without any knowledge about their internal functionality or their configuration (see Section \ref{sec:gui} for more details). Power users can use JedAI-spatial as a library or a Maven dependency, can manually fine-tune the selected methods and can extend it with more algorithms or pipelines according to their needs.

\vspace{3pt}
\noindent
$\bullet$ \textit{Extensibility.} Every algorithm in JedAI-spatial implements the interface of its workflow step, which determines its input and output. Hence, new methods can be seamlessly integrated into JedAI-spatial as long as they implement the corresponding interface so that they are treated like the existing ones. Similarly, new workflow steps can be added as long as they define a new interface specifying their input and output. All additions should implement the \texttt{IDocumentation} interface (see Section \ref{sec:auxiliaryComponents} for more details).

\vspace{3pt}
\noindent
$\bullet$ \textit{Efficiency and scalability.} JedAI-spatial scales well to large datasets both in commodity/stand-alone systems and in computer clusters that run Apache Spark. For more details, refer to the implementation improvements in Section \ref{sec:backEnd} and the experiments in Section \ref{sec:quantAnalysis}.

\section{Back-end: JS-core}
\label{sec:backEnd}

All methods have been re-implemented in JedAI-spatial's common framework, thus minimizing the dependencies to other systems and libraries. For most algorithms, we have incorporated improvements that significantly enhance their original performance. 

\subsection{Serial Algorithms}
\label{sec:serialAlg}

Following all relevant open-source libraries in the literature (i.e., Silk and LIMES), the serial algorithms of JS-core are implemented in Java, which facilitates the deployment of our system (due to its portability), its use as a library (through Maven), its extension by practitioners and researchers (through the public interfaces) as well as 
its maintenance (due to its object-oriented capabilities).


\subsubsection{Budget-agnostic algorithms}
\label{sec:batchSerial}

\begin{figure}[t]
\centering
\includegraphics[width=0.38\textwidth]{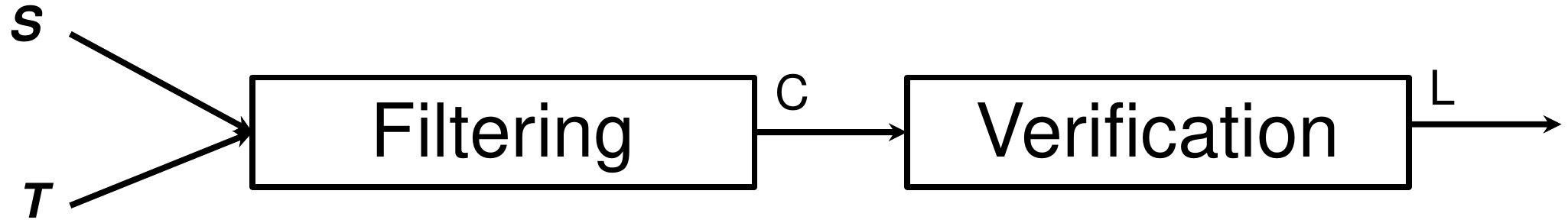}
\vspace{-5pt}
\caption{The pipeline of budget-agnostic algorithms.}
\vspace{-12pt}
\label{fig:fvFramework}
\end{figure}

The methods of this category address Problem \ref{pr:batchGI}. To compute all positive topological relations between the source and the target geometries, $S$ and $T$, respectively, they follow the two-step pipeline in Figure \ref{fig:fvFramework}: initially, the Filtering step indexes the source dataset and, if necessary, the target one, based on the minimum bounding rectangle (\textbf{MBR}) of each geometry -- in Figure \ref{fig:example}, the MBRs are the dotted rectangles surrounding each geometry. The resulting index is used to generate $C$, the set of candidate pairs, which are likely to satisfy at least one topological relation. Next, the Verification step examines every pair in $C$ as long as their MBRs are intersecting. The detected topological relations are added to the set of triples $L$, which is returned as output.

JedAI-spatial organizes these algorithms
into the following three subcategories, based on the type of the index used in Filtering. 

\vspace{3pt}
\noindent
\textbf{Grid-based Algorithms.} The input geometries are indexed by dividing the Earth's surface into cells of the same dimensions. The index is called \textit{Equigrid} and its cells \textit{tiles}. Every geometry is placed into the tiles that intersect its MBR. JedAI-spatial conveys four state-of-the-art algorithms of this type, which differ in the definition and use of the Equigrid during Filtering and Verification.

\vspace{3pt}
\noindent
\textit{RADON \cite{DBLP:conf/aaai/SherifDSN17}.} Filtering loads both input datasets into main memory and defines an Equigrid index by setting the horizontal and vertical dimensions of its tiles equal to the average width and height, respectively, over all geometries.
Verification computes the Intersection Matrix for all candidate pairs \cite{DBLP:conf/semweb/AhmedSN18}, taking special care to avoid the ones repeated across different tiles of the Equigrid.

\vspace{3pt}
\noindent
\textit{GIA.nt \cite{DBLP:conf/www/0001MMK21}.} Filtering loads into main memory only the input dataset with the fewest geometries. The granularity of the Equigrid index is determined by the average dimensions of this dataset. Verification reads the geometries of the other dataset from the disk, one by one. For each geometry $g$, it sets as candidates those with an MBR intersecting the same tiles as $MBR(g)$ and $MBR(g)$ itself. Then, it computes their Intersection Matrix, adding the detected links to $L$.

\vspace{3pt}
\noindent
\textit{Static variants.} Unlike the \underline{dynamic} Equigrid of the above algorithms, whose granularity depends on the input data, \textit{Silk-spatial} \cite{DBLP:conf/www/SmerosK16} employs a \underline{static} Equigrid, whose granularity is predetermined (by the user), independently of the input characteristics. Even though the resulting index might be too fine- or coarse-grained for the input datasets, this approach is based on the idea that the candidate pairs are eventually filtered out if their MBRs are disjoint. To put this approach into practice, JedAI-spatial includes the custom methods \textit{Static RADON} and \textit{Static GIA.nt}, where the index granularity is a-priori defined by the user.

\vspace{3pt}
\noindent
\textit{Implementation improvements.} RADON's implementation is publicly available through LIMES\footnote{\url{https://github.com/dice-group/LIMES}}. However, we re-implemented it in JedAI-spatial so as to significantly improve its performance. First, we reduce the run-time of Filtering by skipping the swapping strategy, which is used to identify the input dataset with the smallest overall volume (this strategy has no impact on Filtering, given that the Equigrid granularity considers both input datasets). Second, we reduce RADON's memory footprint to a significant extent. Instead of a hashmap that stores all examined pairs in main memory to avoid verifying the same candidate pairs more than once, we use the \textbf{reference point technique} \cite{DBLP:conf/icde/DittrichS00}, verifying every candidate pair only in the tile that contains the top left corner of their intersection; as an example, consider the geometries $g_3$ and $g_4$ in Figure \ref{fig:example}: they co-occur in the tiles (4,A), (4,B), (5,A), (5,B), but are verified only in (5,A), where their reference point (black dot) lies. Moreover, unlike the original implementation, which refers to all geometries by their URL (of type \texttt{String}), we use ids for this purpose (of type \texttt{int}). We also use the data structures of the GNU Trove library \cite{friedman2013gnu}, which work with primitive data types (e.g., the 4-bytes \texttt{int} instead of the 16-bytes \texttt{Integer}). The same improvements apply to the methods corresponding to Silk-spatial, namely the static variants. For GIA.nt \cite{DBLP:conf/www/0001MMK21}, we use its open-source implementation\footnote{\url{https://github.com/giantInterlinking/prGIAnt}}, which already involves these optimizations.

\vspace{3pt}
\noindent
\textbf{Partition-based Algorithms.} They rely on a (usually vertical) sweep line that moves across the Earth's surface, stopping at some points. Filtering sorts all input geometries in ascending order of their lower boundary on the horizontal axis, $x_{min}$. Verification is restricted to pairs of source and target geometries whose MBRs simultaneously intersect the sweep line whenever it stops. The process terminates once the sweep line passes over all geometries.

\vspace{3pt}
\noindent
\textit{Plane Sweep \cite{DBLP:conf/sigmod/BrinkhoffKS93}.} This cornerstone algorithm applies the above process to all source and target geometries. Before verifying a pair of geometries, it ensures that they overlap on the y-axis.

\vspace{3pt}
\noindent
\textit{PBSM \cite{DBLP:conf/sigmod/PatelD96}.} This algorithm splits the given geometries into a manually defined number of orthogonal partitions and applies Plane Sweep inside every partition. Filtering defines the partitions, assigns every geometry to all partitions that intersect its MBR and sorts all geometries per partition in ascending $x_{min}$. Verification goes through the partitions and in each of them, it sweeps a vertical line $l$, computing the Intersection Matrix for each pair of geometries that simultaneously intersect $l$ and overlap on the $y$-axis. To avoid repeated verifications of the same geometry pairs across different partitions, it uses the \textit{reference point technique} \cite{DBLP:conf/icde/DittrichS00}.

\vspace{3pt}
\noindent
\textit{Stripe Sweep.} To lower the time complexity of Plane Sweep, this \underline{new} algorithm
sorts only the geometries of the smallest input dataset during Filtering, i.e., the source dataset. These geometries are then partitioned into several vertical stripes, whose length is equal to the average width of the source geometries. Every source geometry is placed in all stripes that intersect its MBR. Verification probes every target geometry $t$ against the stripes and aggregates the \underline{set} of the source geometries that intersect the same stripes with $t$; this way, it gathers the distinct candidate geometries, avoiding redundant verifications. This set is further refined by retaining only the candidate pairs with intersecting MBRs.

\vspace{3pt}
\noindent
\textit{Implementation improvements.} Plane Sweep and PBSM employ a dynamic data structure, called \textit{sweep structure}, which stores in main memory the \underline{active geometries}, i.e., the ones whose MBR intersects the sweep-line in its current position. JedAI-spatial supports two different sweep structures: (i) \textit{List Sweep} maintains one linked list for each input dataset. In every move of the sweep line $l$, the contents of both lists are updated, inserting the geometries with an intersecting MBR and removing the expired ones, i.e., the geometries with $x_{max} < l_x$. (ii) \textit{Striped Sweep} splits the given datasets into $n$ stripes and uses a different List Sweep per stripe. After preliminary experiments, the length of each stripe on the horizontal axis was set to the average width of the source geometries.

Stripe Sweep can use two different data structures for storing the source geometries per stripe: (i) a hash map, which associates every stripe id with the corresponding source geometry ids, and (ii) an STR-Tree \cite{DBLP:conf/icde/LeuteneggerEL97}, which indexes the source geometries in each stripe. The hash map does not ensure the overlap on the y-axis 
before checking the MBR intersection of candidate pairs, unlike the STR-Tree. 
For the implementation of the STR-Tree, we use the optimized implementation provided by the JTS library\footnote{\url{https://locationtech.github.io/jts/}}.

\begin{figure}[t]
\centering
\includegraphics[width=0.47\textwidth]{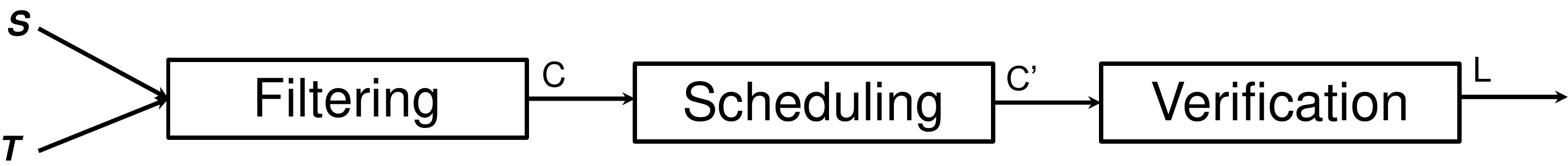}
\vspace{-10pt}
\caption{The pipeline of budget-aware algorithms.}
\vspace{-12pt}
\label{fig:fsvFramework}
\end{figure}

\vspace{3pt}
\noindent
\textbf{Tree-based Algorithms.} As suggested by their name, these algorithms rely on state-of-the-art spatial tree indices. During Filtering, they index the smallest input dataset. During Verification, every geometry $g$ from the other dataset queries the tree index; its candidates are located in the leaf nodes whose MBR intersects with $MBR(g)$. For all candidate geometries with an MBR that intersects $MBR(g)$, the Intersection Matrix is computed.

\vspace{3pt}
\noindent
\textit{R-Tree \cite{DBLP:conf/sigmod/Guttman84}.} In this index, every non-leaf node contains pointers to its child nodes along with an MBR that encloses the span of all the MBRs in its children. Every leaf node contains up to $M$ geometries. 
When an entry is added to a full node, the node is split into two new ones, which are initialized with the two largest geometries. The remaining geometries are added to the node whose MBR expands the least after insertion.

\vspace{3pt}
\noindent
\textit{Quadtree \cite{DBLP:journals/acta/FinkelB74}.} In this index, every non-leaf node has exactly four children, dividing the space into four quadrants: NorthEast (NE), NorthWest (NW), SouthEast (SE) and SouthWest (SW). Again, every node has a maximum capacity $M$. When $M$ is reached, the corresponding cell is split into four new ones, its children.

\vspace{3pt}
\noindent
\textit{CR-Tree \cite{DBLP:conf/sigmod/KimCK01}.} This index compresses the R-Tree so that it leverages the L1 and L2 cache memory of CPUs, which have faster access times. Using the \textit{Quantized Relative Representation of MBR}, it minimizes the size of the MBRs, which dominate the space requirements. CR-Trees are usually wider and smaller than R-Trees, achieving higher time efficiency, while occupying $\sim$60\% less memory.

\vspace{3pt}
\noindent
\textit{Implementation improvements.} For R-Tree and CR-Tree, we leverage the data structures of GNU Trove, which work with primitive data types, offering high time efficiency and low memory footprint. For Quadtree, we use the optimized implementation of the JTS library.

\subsubsection{Budget-aware algorithms}
\label{sec:progressiveSerial}

This category encompasses methods that address Problem \ref{pr:progressiveGI}. 
Their goal is to maximize the number of related geometry pairs that are detected after consuming the available budget $BU$, which determines the maximum number of verifications. To this end, they follow the three-step pipeline in Figure \ref{fig:fsvFramework}. Filtering is identical with that of budget-agnostic methods, producing a set of candidate pairs $C$. Scheduling first refines $C$ by discarding the pairs with non-overlapping MBRs. Then, it defines the processing order of the remaining pairs so that the likely related ones are placed before the unlikely ones. The new set of candidate pairs $C'$ is forwarded to Verification, which carries out their processing and returns the set of detected links, $L$.

The gist of budget-aware algorithms is the combination of Scheduling with Filtering, as Verification remains the same in all cases. Based on the co-occurrence patterns of \underline{grid-based} Filtering, Scheduling assigns a score to every pair of candidates with intersecting MBRs. The higher this score is, the more likely the constituent geometries are to satisfy at least one topological relation. JedAI-spatial offers the following weighting schemes \cite{DBLP:conf/www/0001MMK21,DBLP:conf/tsas/0001MMK22}:
\begin{itemize}
    \item \textit{Co-occurrence Frequency} (\textsf{CF}) measures how many tiles intersect both the MBR of the source and the target geometry.
    \item \textit{Jaccard Similarity } (\textsf{JS}) normalizes \textsf{CF} by the number of tiles intersecting each geometry.
    \item \textit{Pearson's $\chi^2$ test} extends \textsf{CF} by assessing whether the given geometries appear independently in the tiles.
    \item \textit{Minimum Bounding Rectangle Overlap} (\textsf{MBRO}) returns the normalized overlap of the MBRs of the two geometries.
    \item  \textit{Inverse Sum of Points} (\textsf{ISP}) amounts to the inverse sum of boundary points in the two geometries, thus promoting the simpler candidate pairs.
    \item \textit{Composite Weighting Schemes} combine two of the above atomic schemes. The primary scheme determines the processing order of all candidate pairs, while the secondary scheme breaks the ties of the primary one.
\end{itemize}

\noindent
These weighting schemes are leveraged by the following algorithms.

\vspace{3pt}
\noindent
\textit{Progressive GIA.nt \cite{DBLP:conf/www/0001MMK21}.} It applies the same Filtering as its budget-agnostic counterpart and, then, its Scheduling gathers in a priority queue the top-$BU$ weighted candidate pairs.

\vspace{3pt}
\noindent
\textit{Dynamic Progressive GIA.nt \cite{DBLP:conf/tsas/0001MMK22}.} It uses the same Filtering and Scheduling as Progressive GIA.nt. Its Verification, though, does not employ a static processing order. Instead, it updates the processing order of the top-weighted candidate pairs dynamically, as more topologically related pairs are detected: whenever a pair of geometries $(s, t)$ is detected as topologically related, it updates the weight of all top-ranked candidate pairs that include $s$ or $t$, but have not been processed yet. To minimize the cost of updating pair weights, a tree set is used instead of a priority queue. 

\vspace{3pt}
\noindent 
\textit{Local Progressive GIA.nt.} To ensure that every source or target geometry is represented in the $BU$ retained candidate pairs, this \underline{new} algorithm
applies the same Filtering as GIA.nt, but its Scheduling retains a portion of 
candidates per target geometry.
The top-weighted $BU$ candidates are then forwarded to Verification.

\vspace{3pt}
\noindent 
\textit{Geometry-ordered GIA.nt.} This is another \underline{new} progressive algorithm that assumes that the larger the average weight of a geometry is, the more likely it is to be related to its candidates. Thus, it applies the same Filtering as GIA.nt and then, it estimates the average weight per source or target geometry. After sorting the geometries in decreasing average weight, it selects the $BU$ candidates from the top-weighted geometries and sorts them in decreasing weight.

\vspace{3pt}
\noindent
\textit{Iterative Progressive GIA.nt}. Unlike the previous approach, this \underline{new} algorithm goes through the sorted list of geometries iteratively: in every iteration, it examines the next top-weighted pair for the next top-weighted source or target geometry. 

\vspace{3pt}
\noindent
\textit{Progressive RADON \cite{DBLP:conf/www/0001MMK21}.} It applies RADON's Filtering and defines the processing order of the resulting tiles by sorting them in increasing or decreasing number of candidate pairs. Inside every tile, it identifies the non-redundant candidate pairs using the reference point technique. Those with intersecting MBRs, are processed in decreasing score, as determined by the selected weighting scheme. Thus, the pairs that are most likely to satisfy topological relations are processed first inside every tile.

\begin{figure*}[t]
\centering
\includegraphics[width=0.89\textwidth]{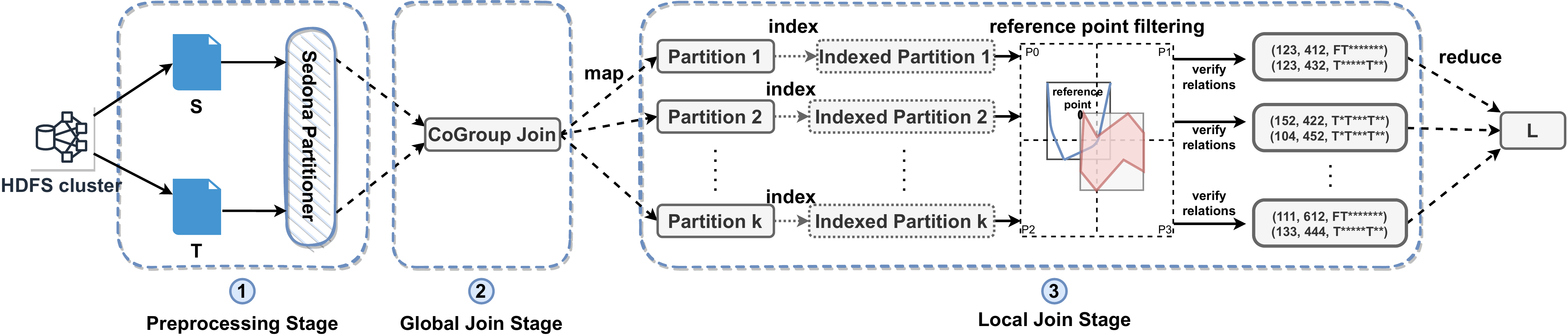}
\vspace{-5pt}
\caption{The three-step pipeline of parallel, budget-agnostic algorithms in JedAI-spatial.}
\vspace{-12pt}
\label{fig:parallelFR}
\end{figure*}

\subsection{Parallel Algorithms}
\label{sec:parallelAlg}

To scale to voluminous datasets, JedAI-spatial exploits the massive parallelization functionalities offered by the state-of-the-art framework of Apache Spark.
JedAI-spatial has aggregated all relevant algorithms in the literature that are crafted for the same framework and the same types of geometries, i.e., LineStrings and Polygons \cite{DBLP:journals/pvldb/PandeyKNK18} (we exclude SIMBA \cite{DBLP:conf/sigmod/XieL0LZG16}, which exclusively applies to points). 

We adapted all algorithms to the pipeline in Figure \ref{fig:parallelFR} so as to reduce the time-consuming Spark shuffles, increasing the overall performance. The pipeline consists of the following three steps:
\begin{enumerate}
    \item The \textit{Preprocessing Stage} reads the source and target datasets from HDFS, 
    transforms them into Spark RDDs and 
    partitions them according to a predetermined approach.
    \item The \textit{Global Join Stage} joins the source and target partitions that 
    are overlapping and assigns every pair of overlapping partitions to a different worker for processing.
    \item In the \textit{Local Join Stage}, each worker interlinks the assigned 
    pairs of overlapping source and target partitions.
\end{enumerate}

Below, we explain how the main parallel interlinking algorithms are adapted to this three-step pipeline.

\subsubsection{Budget-agnostic algorithms} 
\label{sec:baParallel}

JedAI-spatial conveys the following four state-of-the-art parallel approaches:

\vspace{2pt}
\noindent
\textit{GeoSpark \cite{DBLP:conf/gis/YuWS15}.} This algorithm is now part of \textit{Apache Sedona}\footnote{\url{http://sedona.apache.org}}. During Preprocessing, it uses sampling to partition the input data with a KDB-Tree or a Quadtree. The geometries that are not covered by the index are added to an overflow partition. The overlapping source and target partitions are assigned to the workers, during the Global Join Stage. The Local Join Stage verifies the candidate pairs through a nested loop join or indexes the source geometries with an R-Tree or a Quadtree that is then queried by the target ones.

\vspace{3pt}
\noindent
\textit{Spatial Spark \cite{DBLP:conf/icde/YouZG15}.} This algorithm entails two functionalities:

(i) The \textit{broadcast join} supports up to 2GB of source data. During the Preprocessing Stage, the source dataset is indexed by an R-Tree, which is then broadcast as a read-only variable to all workers. Every worker also receives a disjoint partition of the target geometries. The Global Join Stage is skipped. During the Local one, every worker iterates over its target geometries, retrieves the candidate source ones from the R-Tree and verifies those intersecting the target MBR.

(ii) The \textit{partition join} overcomes the size limit of the broadcast join by implementing all three steps of JedAI-spatial's framework. Depending on the user's choice, the Preprocessing Stage indexes the entire input data (in case of a Fixed Grid Partition, whose dimensions, $dim_X \times dim_Y$ are defined by the user) or a sample of the source and target data (in case of Binary Split or Sort Tile Partitions, which use an R-Tree). The Global Join Stage assigns the overlapping source and target partitions to the same worker so that their candidate pairs are verified locally, during the third stage.

\vspace{3pt}
\noindent
\textit{Magellan.}\footnote{\url{https://github.com/harsha2010/magellan}} This algorithm relies on the Z-Order Curves, which define an Equigrid on the Earth's surface during the Preprocessing Stage. The number of tiles in this grid is determined as $2^p$, where $p$ is the precision parameter that is set by the user. Apparently, the higher the precision is, the more fine-grained is the resulting Equigrid index. The Global Join Index sends to the same workers the source and target geometries that intersect the same tiles. During the Local Join Index, every worker checks every candidate pair 
and verifies those with intersecting MBRs.

\vspace{3pt}
\noindent
\textit{Location Spark \cite{DBLP:journals/pvldb/TangYMOA16,DBLP:journals/fdata/TangYMMOA20}.}
Its Preprocessing partitions the source and target datasets using a Grid, R-Tree or Quadtree index. Then, its Query Plan Scheduler performs a skew analysis in order to partition the data as evenly as possible, balancing the workload among the workers. In essence, it repartitions the skewed partitions, which include at least twice as many geometries as the smallest one. After joining the overlapping source and target partitions during the Global Join Stage, a local index is constructed for the source geometries of every worker using an R-Tree, QuadTee or EquiGrid. The Local Join Stage queries the index with the target geometries and verifies the candidate pairs with intersecting MBRs.

\vspace{3pt}
\noindent
\textit{Parallel GIA.nt \cite{DBLP:conf/www/0001MMK21}.} The Preprocessing estimates the average width and height of the source geometries. These dimensions, which are broadcast to all workers, define the Equigrid that partitions both input datasets. The next stage joins the overlapping source and target partitions, while the Local Join Stage creates an Equigrid of the source geometries inside every worker, using the broadcast dimensions. The target geometries query the index to retrieve the candidates with intersecting MBRs, which are then verified. The reference point technique eliminates all repeated verifications.

\vspace{3pt}
\noindent
\textit{Implementation Improvements.} 
The most important enhancement to all parallel algorithms is the use of the reference point technique during the Local Join Stage in order to avoid all repeated verifications. This has replaced GeoSpark's \texttt{groupBy}, Spatial Spark's call to \texttt{distinct}, Magellan's \texttt{dropDuplicates} and Location Spark's call to \texttt{reduceByKey}. All these functions shuffle the output data along the cluster, imposing significant overhead.

We also replaced Magellan's Extended Spark SQL with Spark RDDs for higher efficiency. 
For Location Spark, we removed the Spatial Bloom Filter; it is mainly used for spatial range queries, but in our case, where every geometry is assigned to multiple partitions, it imposes an unnecessary overhead. From GeoSpark, we removed the R-Tree from the indexes supported by the Global Join Stage: due to sampling, its overflow bucket typically contains geometries from the entire input datasets and, thus, it is not disjoint from the rest of the partitions; as a result, the reference point technique cannot be used to eliminate redudant verifications. For Parallel GIA.nt, we use the implementation from {\small \url{https://github.com/giantInterlinking/prGIAnt}}.

\subsubsection{Budget-aware algorithms} JedAI-spatial parallelizes all serial budget-aware algorithms described in Section \ref{sec:progressiveSerial}. The Preprocessing and Global Join Stage are identical with Parallel GIA.nt. Then, the overall budget $BU$ is split among the partitions assigned to every worker based on the portion of candidate pairs it involves. The Local Join Stage applies the budget-aware algorithm to the data assigned to every worker, using the corresponding local budget.
\vspace{-5pt}
\subsection{Auxiliary Components}
\label{sec:auxiliaryComponents}

We now describe the rest of the components in Figure \ref{fig:architecture}, which play an important role in the characteristics offered by JedAI-spatial.

\vspace{3pt}
\noindent
\textit{Data Model.} This component implements the classes and the data structures that lie at the core of JedAI-spatial. The cornerstone is the \texttt{GeometryProfile} class, which supports all heterogeneous data formats mentioned in Section \ref{sec:architecture}. This is accomplished by representing every geometry as a set of name-value pairs, which capture the textual information about an entity, coupled with a \texttt{Geometry} object of the JTS library that is accompanied by its MBR and the method for computing an Intersection Matrix. This simple, yet versatile \texttt{GeometryProfile} class also facilitates the visualization and inspection of input data through JS-gui.

\vspace{3pt}
\noindent
\textit{Documentation.} This component essentially corresponds to a Java interface that is implemented by all algorithms. The interface conveys methods providing textual information about the most important aspects of each algorithm: its name, a summary of its functionality, the name of every configuration parameter, a short description of every parameter, the domain of every parameter (i.e., its default, minimum and maximum values) as well as the configuration of the current algorithm instantiation. This information is provided to the user through tooltips in the Web application and through the help option of the command line interface.

\vspace{3pt}
\noindent
\textit{Parameter-configuration.} JedAI-spatial facilitates the fine-tuning of any supported algorithm, because a poor parameterization invariably leads to poor performance. 
Three modes are supported: 
\begin{enumerate}
    \item \textit{Default configuration} a-priori sets all parameters of each algorithm to values that empirically achieve reasonable performance across different datasets. This mode allows lay users to apply the desired pipeline to their data simply by choosing among the available methods.
    \item \textit{Manual configuration} enables power users to fine-tune an algorithm themselves, based on their own experience or on the information provided by the \texttt{Documentation} component.
    \item \textit{Grid search} automatically identifies the optimal configuration through a brute-force approach that tries all reasonable values in the domain of each parameter. In the case of budget-agnostic pipelines, the parameterization that minimizes the run-time is selected as the optimal one. For budget-aware pipelines, the optimal parameterization is the one maximizing Progressive Geometry Recall.
\end{enumerate}
\section{Front-end: JS-gui}
\label{sec:gui}

\begin{figure}[t]
\centering
\includegraphics[width=0.5\textwidth]{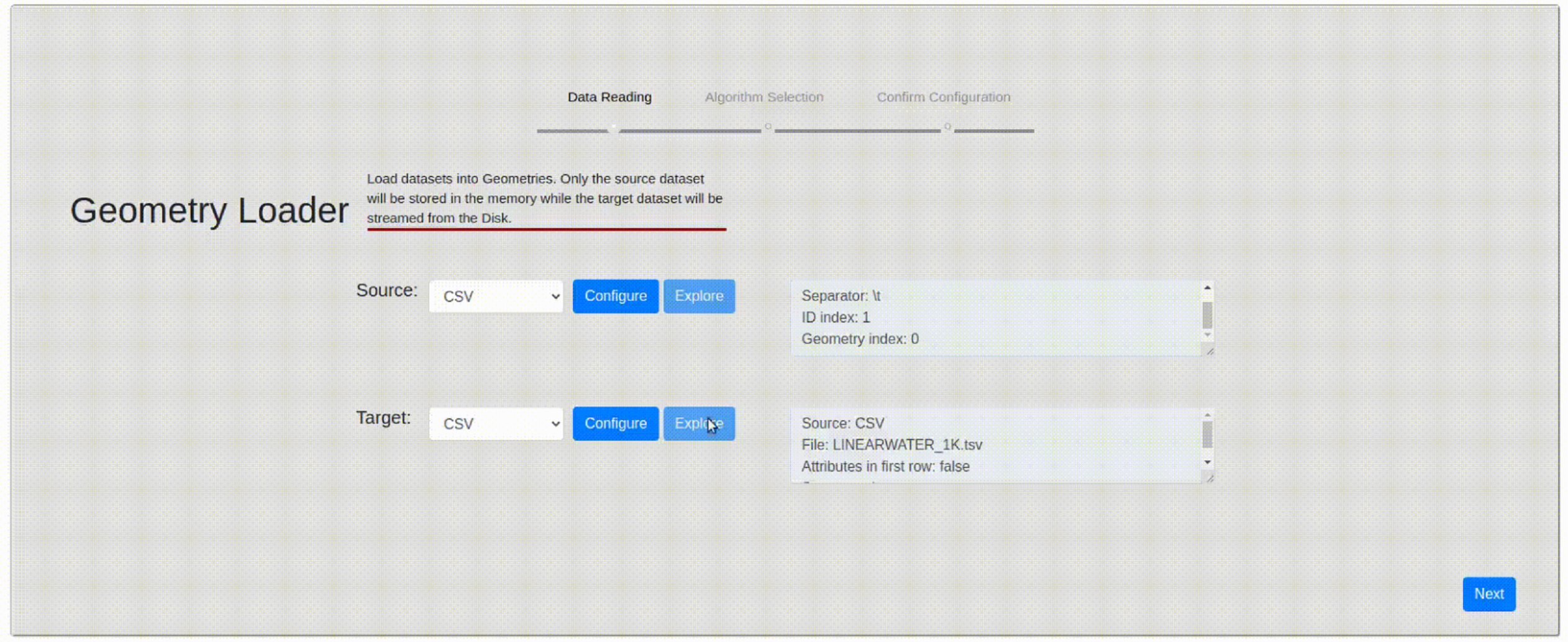}\\
\vspace{-5pt}
{\center (a) The data reading screen.}\\
\includegraphics[width=0.5\textwidth]{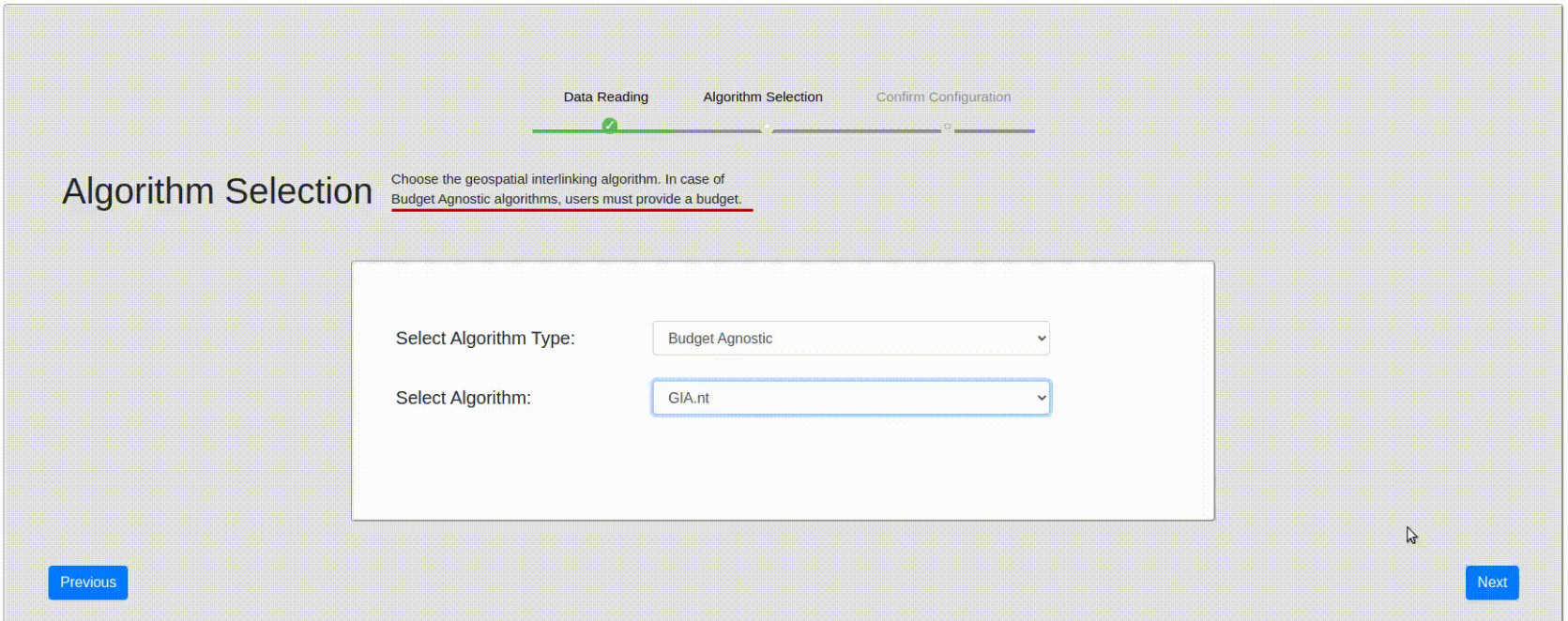}\\
\vspace{-5pt}
{\center (b) The algorithm selection screen.}\\
\includegraphics[width=0.5\textwidth]{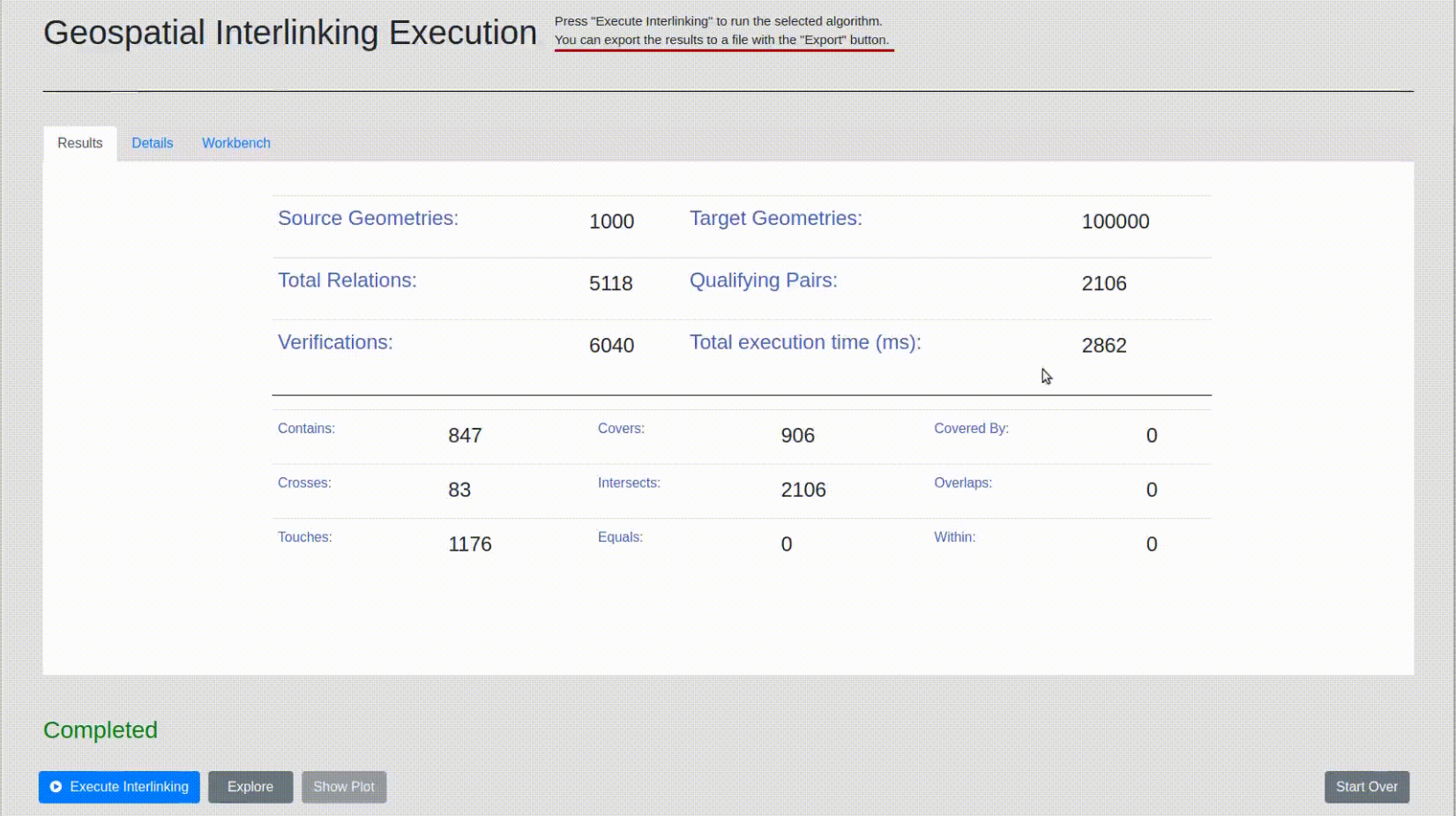}\\
\vspace{-5pt}
{\center (c) The results screen.}
\vspace{-8pt}
\caption{Screenshots of JS-gui.}
\vspace{-15pt}
\label{fig:jsgui}
\end{figure}

JedAI-spatial support users of any experience level, offering two wizard-like user interfaces that simplify its use to a great extent: 
\begin{enumerate}
    \item The command line interface. JS-core produces an executable jar, which when run, guides users in applying the desired pipeline to their data.
    \item The Web application interface. JS-gui is available as a Docker image, which, when deployed, runs a Web application that seamlessly supports both serial and parallel execution -- the latter relies on \textit{Apache Livy} (\url{https://livy.apache.org}). 
\end{enumerate}

In both cases, users do not need to write code in order to interlink their spatial data. Initially, the data reading screen in Figure \ref{fig:jsgui}(a) requires them to provide the paths of their dataset and the corresponding reading parameters (e.g., the separator character in CSV files). Next, the algorithm selection screen in Figure \ref{fig:jsgui}(b) allows users to select the desired pipeline, i.e., serial or parallel, budget-aware or budget-agnostic, as well as the desired algorithm among the available ones for the selected pipeline. This applies even to the parallel pipelines that run on top of Apache Spark. In both interfaces, users can also inspect the input data and store the detected links to a specific~path. 

Special care has also been taken to facilitate the comparison between the available algorithms, regardless of their space tiling and budget-awareness category. JedAI-spatial acts as a \underline{workbench}, encompassing a results screen, shown in \ref{fig:jsgui}(c), which summarizes the performance of the latest runs with respect to the effectiveness measures (i.e., recall, precision, f-measure and progressive geometry recall, if applicable) as well as the efficiency ones (i.e., the run-time in total and in every step of the selected pipeline). This workbench functionality also allows for examining the impact of configuration parameters on the performance of a particular algorithm (e.g., by changing the granularity of the grid index).

\section{Applications on Web Data}
\label{sec:applications}

The plethora of works on Geospatial Interlinking \cite{DBLP:conf/semweb/JentzschIB10, DBLP:conf/semweb/Ngomo13, DBLP:conf/aaai/SherifDSN17, DBLP:conf/semweb/AhmedSN18, DBLP:conf/sigmod/SantipantakisGD19, DBLP:books/sp/20/SantipantakisDVV20, DBLP:conf/www/0001MMK21,DBLP:conf/semweb/SavetaFFN18} stems from its wide range of practical applications. To this attests the special track on Spatial Link Discovery that is part of the Ontology Alignment Evaluation Initiative during the last few years \cite{DBLP:conf/semweb/PourAAAFFFHHHHI21,DBLP:conf/semweb/PourAAFFHHJJKKL20,DBLP:conf/semweb/AlgergawyFFFHHJ19}. Below, we provide concrete examples of Web applications that can benefit from JedAI-spatial.

GeoSensor \cite{DBLP:conf/www/Argyriou0STPGAL18} is a Web application that integrates satellite images from the Copernicus programme with trends extracted from news agencies and Twitter\footnote{\url{https://twitter.com}}. This is achieved through the geospatial information of both sources, namely the geometry surrounding every satellite image and the geometry corresponding to the areas mentioned in news articles and tweets. To accelerate their interlinking, any algorithm supported by JedAI-spatial could be used. 

Our system can also be used for enriching the GeoLink knowledge graph \cite{doi:10.1080/20964471.2018.1469291}, which integrates the most prominent geospatial repositories in the United States. The knowledge graph is available through a SPARQL endpoint and, thus, JedAI-spatial can directly read its geometries and apply any interlinking algorithm in order to increase the connections and links between its data sources.

Another application that could benefit from JedAI-spatial is GeoLinkedData.es \cite{DBLP:conf/gis/BlazquezVSLCG10}. Its goal is to publish and interlink various data collections from Spain's National Geographic Institute and
National Statistic Institute. This initiative was carried out in the context of EU's INSPIRE knowledge base\footnote{\url{https://inspire.ec.europa.eu}}, which aims to build a public infrastructure for geospatial information in Europe. JedAI-spatial is ideal for efficiently interlinking all data sources in INSPIRE.

\begin{table*}[t]\centering
	\caption{The dataset pairs used in our experiments. The number of topologically related pairs amounts to \#\texttt{Intersects}.}
	\vspace{-10pt}
    \begin{tabular}{ | l | r | r  |r  |r  | r  |r  |}
		\cline{2-7}
		\multicolumn{1}{c|}{}&
		\multicolumn{1}{c|}{$\mathbf{D_{1}}$} &
		\multicolumn{1}{c|}{$\mathbf{D_{2}}$} &
		\multicolumn{1}{c|}{$\mathbf{D_{3}}$} &
		\multicolumn{1}{c|}{$\mathbf{D_{4}}$} &
		\multicolumn{1}{c|}{$\mathbf{D_{5}}$} &
		\multicolumn{1}{c|}{$\mathbf{D_{6}}$}
		\\
		\hline
        \hline
        Source Dataset & AREAWATER & AREAWATER & Lakes & Parks & ROADS & Roads\\
        Target Dataset & LINEARWATER & ROADS & Parks & Roads & EDGES & Buildings\\
        \#Source Geom. & 2,292,766 & 2,292,766 & 8,326,942 & 9,831,432 & 19,592,688 & 72,339,926\\
        \#Target Geom.s & 5,838,339 & 19,592,688 & 9,831,432 & 72,339,926 & 70,380,191 & 114,796,567\\
        \hline
        \hline
        Cartesian Product & 1.34 $\cdot$ 10$^{13}$ & 4.49 $\cdot$ 10$^{13}$  & 8.19 $\cdot$ 10$^{13}$ & 7.11 $\cdot$ 10$^{14}$ & 1.38 $\cdot$ 10$^{15}$ & 8.30 $\cdot$ 10$^{15}$\\
        Candidate Pairs & 6,310,640 & 15,729,319 & 19,595,036 & 67,336,808 & 430,597,631 & 257,075,645 \\
        \hline
        \hline
        \#\texttt{Contains} & 806,158 & 3,792  & 267,457 & 5,147,704 & 53,758,453 & 274,953  \\
        \#\texttt{CoveredBy} & 0 & 0  & 1,944,207  & 47,253 & 12,218,868 & 82,828  \\
        \#\texttt{Covers} & 832,843 & 4,692  & 267,713  & 5,284,672 & 53,758,453 & 274,966  \\
        \#\texttt{Crosses} & 40,489 & 106,823  & 217,198  & 5,700,257 & 6,769 & 313,566 \\
        \#\texttt{Equals} & 0 & 0  & 61,712 & 2,047 & 12,218,868 & 18,909\\
        \#\texttt{Intersects} & 2,401,396 & 199,122  & 3,841,922 & 12,145,630 & 163,982,138 & 1,037,153 \\
        \#\texttt{Overlaps} & 0 & 0  & 488,814 & 42,331 & 73 & 54,810 \\
        \#\texttt{Touches} & 1,554,749 & 88,507  & 986,522 & 1,210,230 & 110,216,843 & 331,166 \\
        \#\texttt{Within} & 0 & 0 & 1,943,643 & 47,155 & 12,218,868 & 81,567 \\
        \hline
        \hline
        Total Topological Relations & 5,635,635 & 402,936  & 10,019,188 & 29,627,279 & 418,379,333 & 2,481,027\\
        \hline
	\end{tabular}
	\label{tb:spatialDatasets}
	\vspace{-10pt}
\end{table*}

Another ideal use case for JedAI-spatial is the LinkedGeoData project \footnote{\url{http://linkedgeodata.org}}, which creates a large spatial knowledge base with global coverage by interlinking OpenStreetMap with other, global data sources, like DBPedia 
and Geonames
\cite{DBLP:journals/semweb/StadlerLHA12}. Due to the extreme volume of data, JedAI-spatial's massively parallel algorithms, especially the budget-aware ones, pose the best option for minimizing the cost of computing all DE9IM topological relations.

Finally, another task that could benefit from JedAI-spatial's capabilities is the dataset recommendation for data interlinking \cite{DBLP:conf/esws/EllefiBDT16,DBLP:conf/icwe/LemeLNCD13}. In this task, the input comprises a source dataset $S$ along with a set of candidate datasets and the goal is to identify those candidates that include entities related to those in $S$. Solutions to this task leverage evidence from: (i) the content similarity in the textual objects of entity descriptions \cite{DBLP:conf/www/Nikolovd11}, (ii) the structural similarity in the schemata and the ontologies of each dataset \cite{DBLP:conf/semweb/EmaldiCL15}, and (iii) the existing links that connect entities across different datasets \cite{DBLP:journals/jdiq/MountantonakisT18,DBLP:conf/icwe/LemeLNCD13}. Apparently, JedAI-spatial is only suitable to datasets with geometric entities. Its budget-aware algorithms can be used for quickly identifying topological relations between geometries of different datasets, even if these datasets are initially disjoint, thus boosting the performance of the third type of dataset recommendation methods.
\vspace{-10pt}

\section{Quantitative Analysis}
\label{sec:quantAnalysis}



\vspace{3pt}
\noindent
\textbf{Experimental Setup.} 
All experiments were carried out on a server with Intel Xeon E5-4603 v2 @ 2.20GHz, 32 cores (16 physical), 4 NUMA nodes and 128GB RAM. The serial methods are implemented in Java 15 and the parallel ones in Scala 2.11.12 and Apache Spark 2.4.4. For each time measurement, we performed 5 repetitions and took the average.

In our experiments, we used the same datasets of \cite{DBLP:conf/www/0001MMK21,DBLP:conf/tsas/0001MMK22}, which are popular in the literature \cite{DBLP:conf/icde/EldawyM15,DBLP:conf/gis/TsitsigkosBMT19}.
They comprise real data from two sources: (i) the US Census Bureau TIGER files, i.e., USA’s Area Hydrography (AREAWATER), Linear Hydrography (LINEARWATER), roads (ROADS) and edges (EDGES), and (ii) OpenStreeMap, i.e., lakes (Lakes), parks (Parks), roads (Roads) and buildings (Buildings) from the whole world. These datasets are combined into the six pairs, $D_1$-$D_6$, that are reported in Table \ref{tb:spatialDatasets}.


\vspace{3pt}
\noindent
\textbf{Serial budget-agnostic processing.} 
We now investigate the relative performance of the budget-agnostic algorithms that are presented in Section \ref{sec:batchSerial}. Given that these methods produce the same results, detecting all topological relations reported in Table \ref{tb:spatialDatasets}, we assess their relative performance with respect to their efficiency. That is, we estimate the time required to complete each step in the pipeline of Figure \ref{fig:fvFramework}: (i) the \textit{filtering time}, $\mathbf{t_f}$, and  (ii) the \textit{verification time}, $\mathbf{t_v}$.

First, we perform 
a scalability analysis using $D_1$.
We split this dataset into 10 subsets of increasing size, from 10\% of source and target geometries to 100\% with a step of 10\%. Special care was taken to ensure that the number of related pairs increases in proportion to the size of the input data.  

\begin{figure*}[t]
\centering
\includegraphics[width=0.325\textwidth]{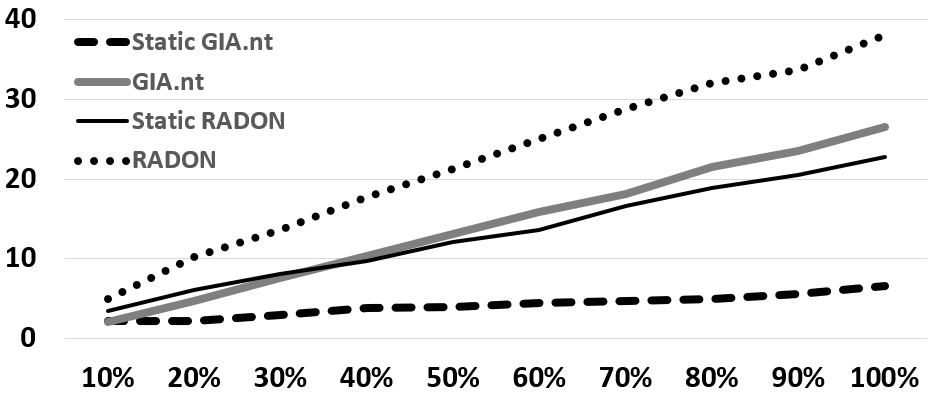}
\includegraphics[width=0.325\textwidth]{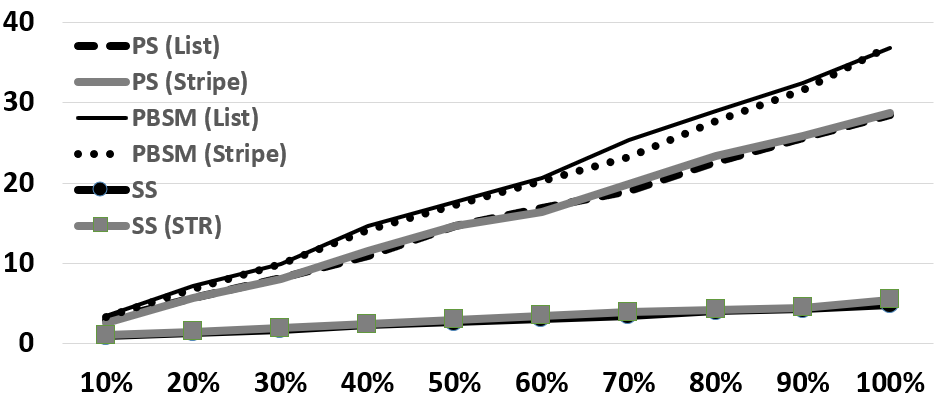}
\includegraphics[width=0.325\textwidth]{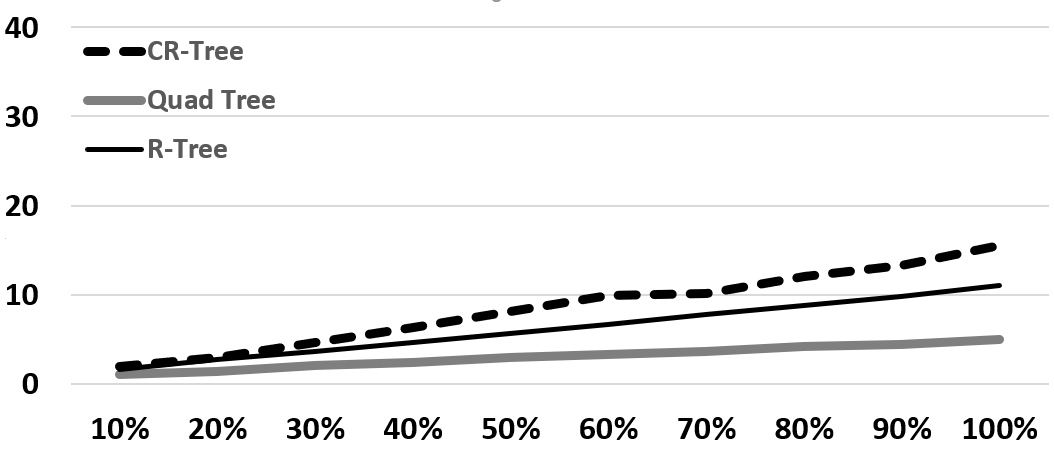}
\vspace{-8pt}
\caption{Scalability analysis of the serial budget-agnostic algorithms with respect to their Filtering time (in seconds).}
\vspace{-5pt}
\label{fig:filteringScalability}
\end{figure*}

\begin{figure*}[t]
\centering
\includegraphics[width=0.325\textwidth]{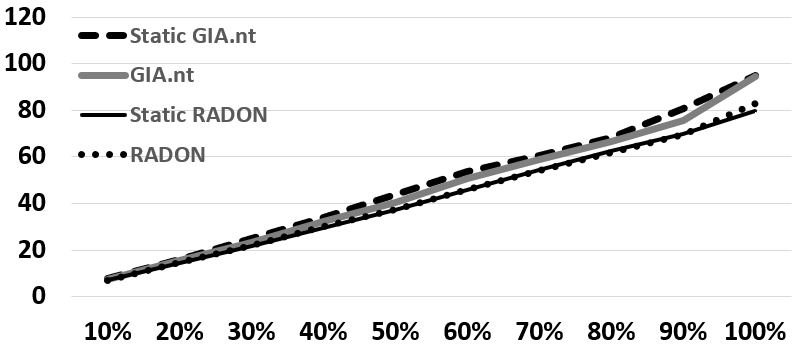}
\includegraphics[width=0.325\textwidth]{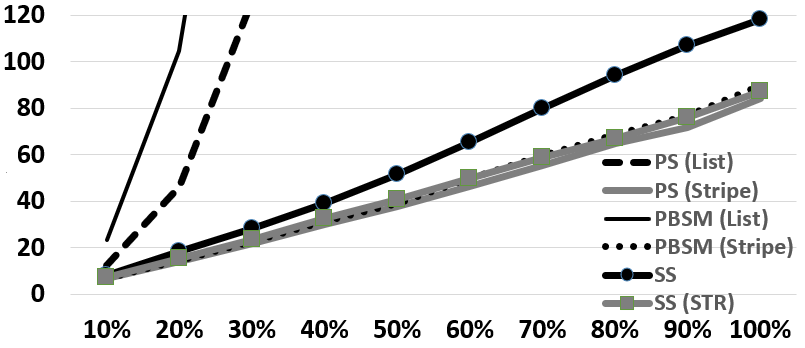}
\includegraphics[width=0.325\textwidth]{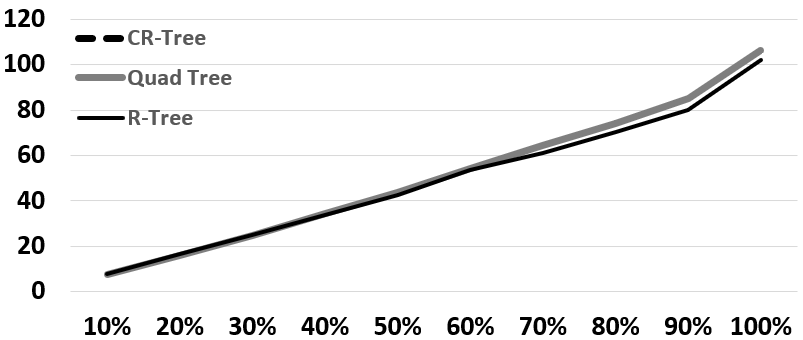}
\vspace{-8pt}
\caption{Scalability analysis of the serial budget-agnostic algorithms with respect to their Verification time (in minutes).}
\vspace{-5pt}
\label{fig:verificationScalability}
\end{figure*}

The resulting filtering and verification times appear in Figures \ref{fig:filteringScalability} and \ref{fig:verificationScalability}, respectively. Each figure encompasses a separate diagram for each algorithm category, but all diagrams use the same scale in order to facilitate the comparisons between the three categories.



\begin{figure*}[t]
\centering
\includegraphics[width=0.49\textwidth]{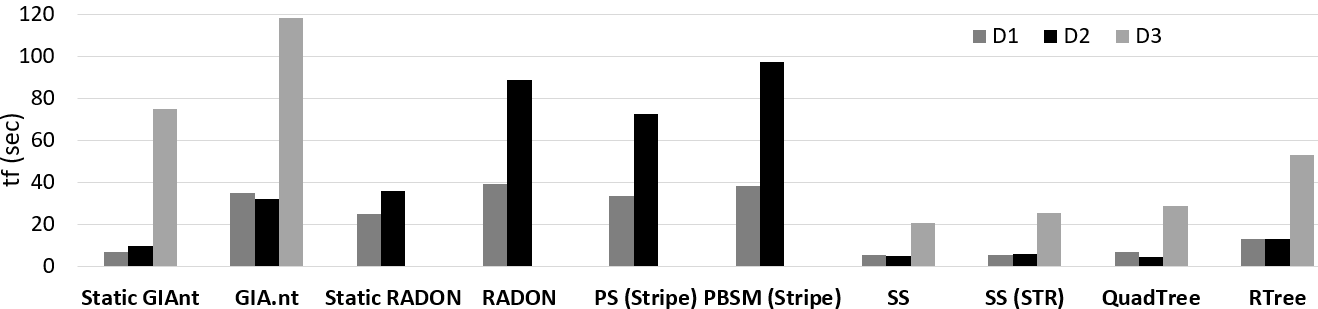}
\includegraphics[width=0.49\textwidth]{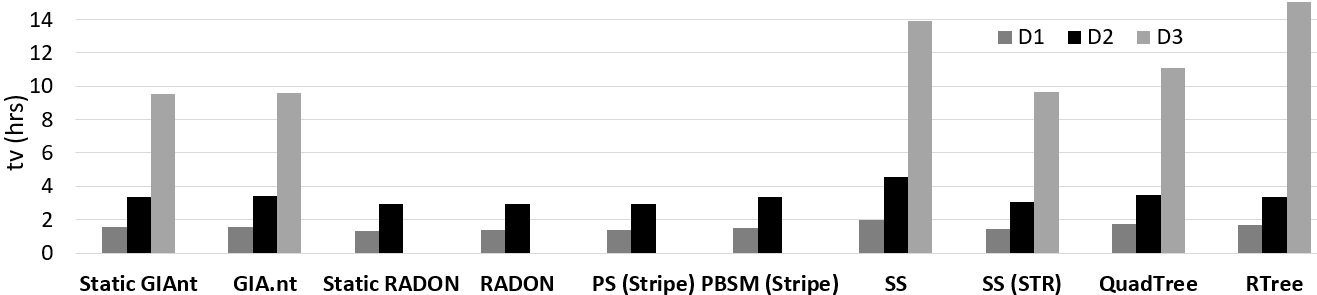}
\vspace{-5pt}
\caption{Filtering time (left) and verification time (right) per serial, budget-agnostic algorithm over $D_1$-$D_3$.}
\vspace{-12pt}
\label{fig:d1d3}
\end{figure*}


Starting with Figure \ref{fig:filteringScalability}, we observe that for all algorithms, the Filtering step is completed within a few seconds, even when processing the entire $D_1$. The reason is that Filtering constitutes a quick process that considers exclusively the MBR of the input geometries, thus disregarding their actual complexity. Yet, it manages to reduce the number of candidates by several orders of magnitude, as shown in Table \ref{tb:spatialDatasets} (Cartesian Product vs Candidate Pairs).

As expected, the algorithms that consider only the source dataset when building their index are much faster than those iterating over both input datasets. The former category includes, (Static) GIA.nt, Stripe Sweep and the tree-based algorithms. All these algorithms scale sublinearly with the increase of the input data: from 10\% to 100\%, their $t_f$ raises by 5 (Stripe Sweep) to 8 (CR-Tree) times. The rest of the algorithms scale linearly with the increase of the input data. We also observe that the static variants of the grid-based algorithms are significantly faster, as they save the cost of deriving the index granularity from the characteristics of the input datasets -- they merely index them. Finally, we notice that the filtering time of each partition-based algorithm is practically stable, regardless of the underlying data structure (List Sweep or Striped Sweep for Plane Sweep and PBSM, hash map or STR-Tree for Stripe Sweep). Overall, \textit{Quadtree and Stripe Sweep exhibit the fastest Filtering.}

Regarding the Verification time, we notice in Figure \ref{fig:verificationScalability} that it constitutes the bottleneck of Geospatial Interlinking, being two orders of magnitude larger than Filtering time. This should be attributed to the complexity of the input geometries, which determines the cost of calculating each Intersection Matrix. We also observe that the algorithms with the fast, source-based Filtering are now slower, because they read the target geometries from the disk, one by one. This overhead is included in their $t_v$, increasing linearly with the size of the input data, hence the larger deviations over larger subsets. \textit{The algorithms (Static) RADON, Plane Sweep and PBSM are faster (in this order)}, because they a-priori load the target geometries into main memory. Note, though, that Plane Sweep and PBSM are by far the slowest algorithms when using a Linked List to maintain their active geometries, due to the high cost of its operations. The performance of both algorithms is significantly improved when using Stripes to reduce the maintenance overhead. The situation is even worse for CR-Tree, which is excluded from Figure \ref{fig:verificationScalability}, because its run-time over the smallest subset is 235 minutes, exceeding the time required by most other algorithms even for the largest subset. The reason is the high cost of retrieving the candidates for every target geometry, due to the compression of MBRs. Finally, we should stress that (Static) RADON is faster than (Static) GIA.nt by 5\% to 12\%, which is in contrast with their relative performance in \cite{DBLP:conf/www/0001MMK21}, due to the significant impact of the implementation improvements we have incorporated in JedAI-spatial. \textit{Yet, GIA.nt involves the fastest verification among the algorithms that read the target geometries from the disk, with Stripe Sweep being slightly slower.}

We now compare the same algorithms with respect to their run-time over $D_1$ and $D_3$. Their filtering times (in seconds) and the verification ones (in hours) are reported in Figure \ref{fig:d1d3}, on the left and right respectively. We exclude datasets $D_4$ and $D_6$, because of their long-lasting execution time. We also exclude CR-Tree as well as Plane Sweep and PBSM in combination with a single Linked List per dataset, due to their poor scalability.

We observe that (Static) RADON, Plane Sweep and PBSM cannot process $D_3$, due to insufficient memory -- they need to load both the source and the target datasets in main memory, but this is not possible with the available 128GB of RAM. 

We also verify some patterns of the scalability analysis. For example, the static grid-based algorithms variants have a faster filtering time than their dynamic counterparts, RADON and GIA.nt, because they do not go through the input geometries when determining the dimensions of their Equigrid. Their verification time, though, is identical with the dynamic grid-based algorithms.

Regarding the remaining algorithms, PBSM's Filtering is significantly slower than that of Plane Sweep, because the latter sorts the input geometries just once. Instead, PBSM sorts the input geometries inside every tile. Due to their coarse granularity, its tiles contain a large number of geometries, thus resulting in an overall computational cost that is higher than that of Plane Sweep. On the other extreme lies Strip Sweep, which has the simplest filtering phase, yielding consistently the lowest filtering time among all methods, followed by Quadtree in close distance. R-Tree involves the next fastest Filtering, as it is outperformed only by Static GIA.nt.

As regards the verification time, Static RADON and RADON are consistently the fastest algorithms, followed in close distance by Plane Sweep. PBSM is a bit slower, because its coarse-grained tiles involve a large number of redundant candidate pairs that are filtered out with the reference point technique. Among the algorithms that index only the source dataset, Stripe Sweep with STR is the fastest one, being slower than RADON by 7.1\% and 5.5\% over $D_1$ and $D_2$, respectively. This difference corresponds to the overhead of reading the target dataset from the disk. (Static) GIA.nt is slower than RADON by $\sim$16\% in all cases, because it creates a much larger number of fine-grained tiles compared to Stripe Sweeep. As a result, GIA.nt examines the content of many more tiles while answering each query (i.e., target geometry). 

High query overhead is also exhibited by Stripe Sweep in combination with a HashMap. Unlike its combination with STR-Tree,  this variant of Stripe Sweep does not filter the candidate pairs with respect to the vertical axis. As a result, its query response time and, thus, its overall verification time increase substantially as we move from the smallest dataset ($D_1$) to the largest one ($D_3$). 

Finally, the tree-based algorithms Quadtree and RTree exhibit a different behavior. For $D_1$ and $D_2$, Quadtree and RTree perform almost as well as GIA.nt. Over $D_3$, though, Quadtree is 16\% slower than GIA.nt, whereas R-Tree is the slowest algorithm by far, with its $t_v$ (57 hours) exceeding the scale of the vertical axis. For both algorithms, this is caused by their sensitivity to parameter configuration and because $D_3$ has a large number of overlapping MBRs. As a result, the range search for candidate geometries visits numerous subtrees recursively, yielding a high time complexity that deviates from the average one, which is discussed in Section \ref{sec:qualitativeAnalysis}.

\textit{To conclude, the most robust serial, budget-agnostic algorithms are (Static) GIA.nt and Strip Sweep STR. Their key characteristic is their simple, but effective filtering phase, which scales sub-linearly with the size of the input data. Their index also allows for faster verification, unlike methods like Quadtree and R-Tree, whose verification time is heavily affected by their parameter configuration and the data characteristics (e.g., overlapping partitions in the spatial index). Finally, RADON, Plane Sweep and PBSM are only suitable for datasets small enough to fit into main memory. In these cases, they provide competitive run-times, especially for the Verification phase, as they save the cost of loading the target dataset from the disk on-the-fly.}

\begin{figure}[t]
\centering
\includegraphics[width=0.44\textwidth]{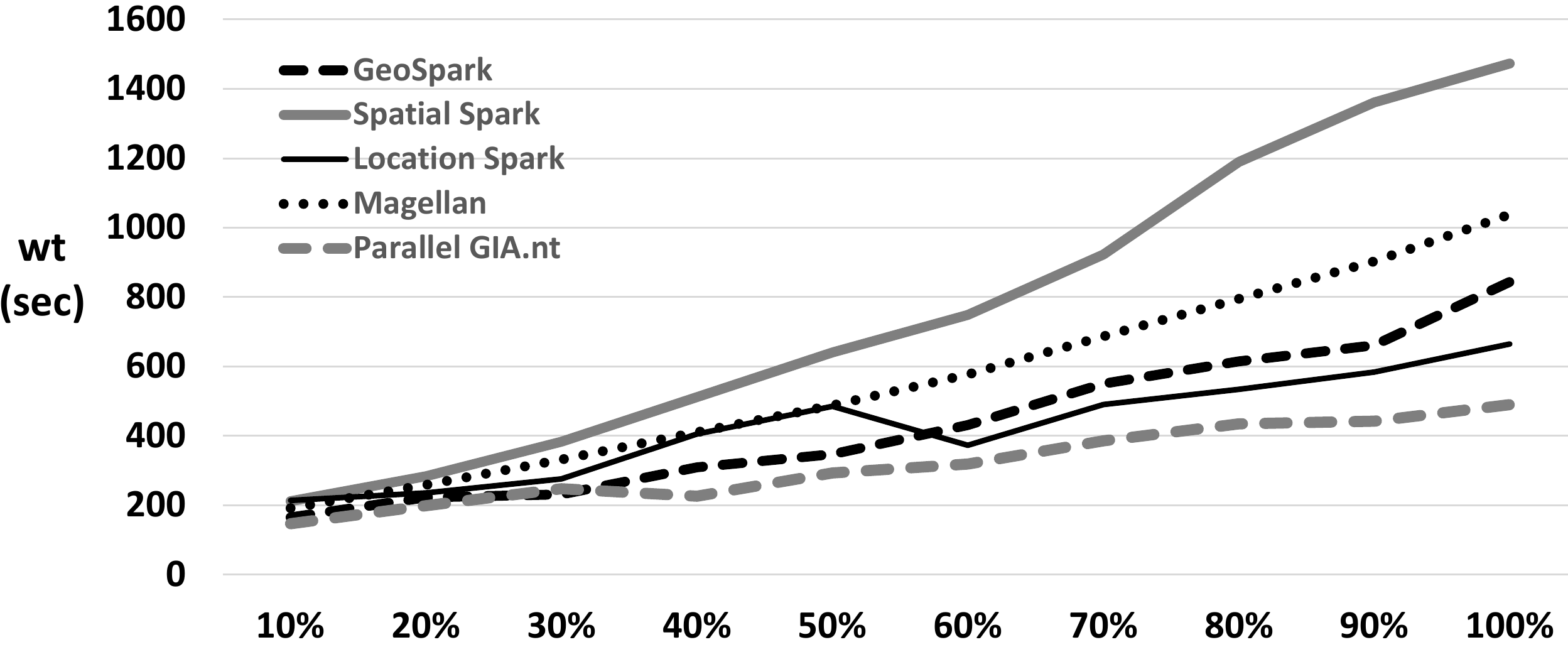}
\vspace{-10pt}
\caption{{\small Scalability of parallel, budget-agnostic algorithms.}}
\vspace{-14pt}
\label{fig:pgr}
\end{figure}

\vspace{3pt}
\noindent
\textbf{Parallel budget-agnostic processing.} We now assess the relative run-time of the budget-agnostic parallel algorithms that are presented in Section \ref{sec:baParallel}. First, we perform a scalability analysis using the same 10 subsets of $D_1$ as in the scalability analysis of serial algorithms. After preliminary experiments, we fine-tune the considered algorithms as follows: GeoSpark (Apache Sedona) is coupled with KDB-Tree partitioning and local indexing with R-Tree, Spatial Spark with a 512x512 Fixed Grid Partitioning and Location Spark with Quadtree partitioning and local indexing with R-Tree. For Magellan, we set precision to 20, while for Parallel GIA.nt, we interchanged the source dataset with the target one.

Figure \ref{fig:pgr} reports the corresponding overall wall-clock times (in seconds). We observe that \textit{Parallel GIA.nt is consistently the fastest algorithm, with Location Spark following in close distance over the largest subsets}, where its skew analysis bears fruit, \textit{leaving GeoSpark in the third place}. These three algorithms require less than half the overall run-time of Spatial Spark, with Magellan lying in the middle of these two extremes. All algorithms scale sublinearly with the size of the input data: from 10\% to 100\%, their run-time increases by 3 (Location Spark, Parallel GIA.nt) to 7 (Spatial Spark) times. 

It is worth stressing at this point that all algorithms are significantly faster than the serial ones, especially over the larger subsets, where the overhead of Apache Spark pays off: for the entire $D_1$, the slowest parallel algorithm (Spatial Spark) is 3.2 times faster than the best serialized algorithm~(RADON).

\begin{figure}[t]
\centering
\includegraphics[width=0.4\textwidth]{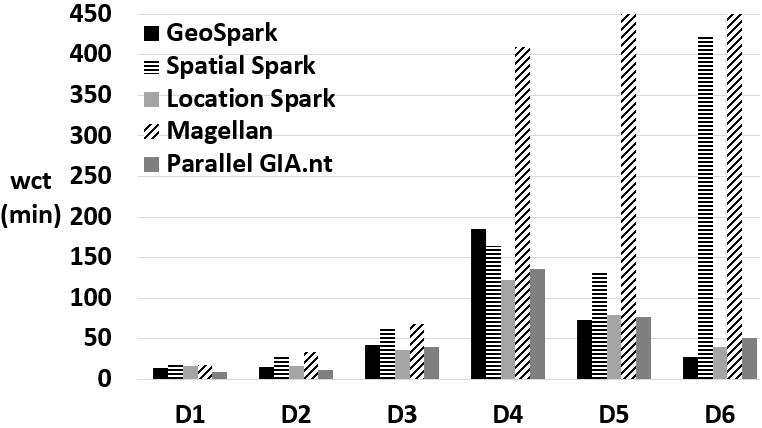}
\vspace{-10pt}
\caption{Performance of the parallel, budget-agnostic algorithms over all dataset pairs in Table \ref{tb:spatialDatasets}.}
\vspace{-14pt}
\label{fig:D1D6parallel}
\end{figure}

Next, we investigate the relative wall clock time of all parallel algorithms over all dataset pairs in Table \ref{tb:spatialDatasets}. After preliminary experiments, we applied the same configurations as in the scalability analysis to all datasets. The only exceptions are Spatial Spark, which is now combined with 512x512 Sort Tile Partitioning, and Parallel GIA.nt, which uses its default configuration, setting the smallest dataset as the source one. The results appear in Figure \ref{fig:D1D6parallel}. Note that for $D_6$, Magellan’s execution was terminated after 24 hours.

We observe similar patterns as in the scalability analysis. Magellan is consistently the slowest approach, with its run-time increasing disproportionately from $D_4$ on. The reason is that its Preprocessing Stage assigns geometries in significantly more partitions than the rest of the algorithms, thus yielding a very high overhead for the two subsequent phases of the parallel framework (see Figure \ref{fig:parallelFR}). The second slowest approach is Spatial Spark, because its sampling-based Sort Tile partitioning scales poorly to large datasets, especially $D_5$ and $D_6$. The rest of the algorithms are yield similar run-times, with each one being the fastest in two datasets: Parallel GIA.nt in $D_1$ and $D_2$, Location Spark in $D_3$ and $D_4$ and GeoSpark in $D_5$ and $D_6$. Remarkably, GeoSpark's execution time over $D_6$ is 29\% lower than the second best, Location Spark. This is counter-intuitive, since the skew analysis of Location Spark is expected to pay off in the largest datasets. In practice, though, the KDB-Tree partitioning of GeoSpark distributes the input data more effectively than the Quadtree of Location Spark over $D_5$ and $D_6$.

\begin{figure*}[t]
\centering
\includegraphics[width=0.7\textwidth]{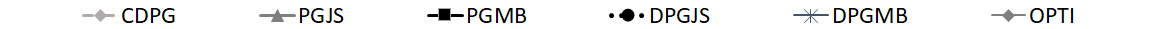}\\
\includegraphics[width=0.97\textwidth]{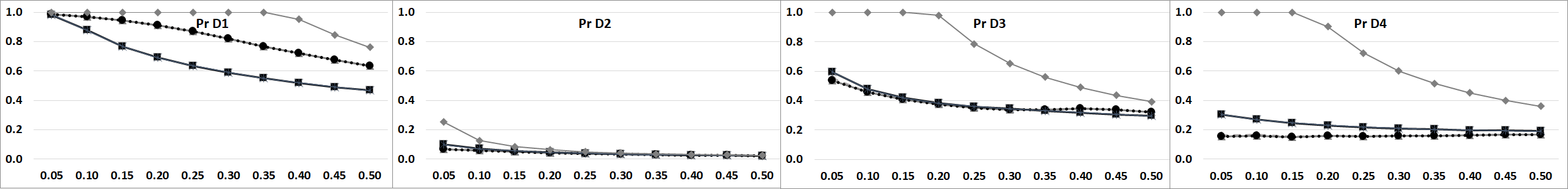}
\includegraphics[width=0.97\textwidth]{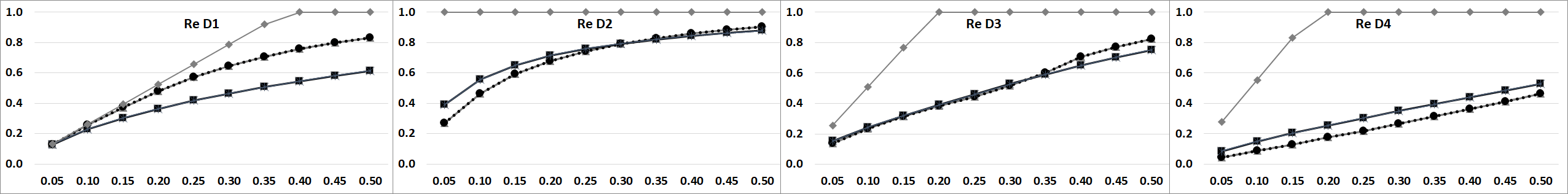}
\includegraphics[width=0.97\textwidth]{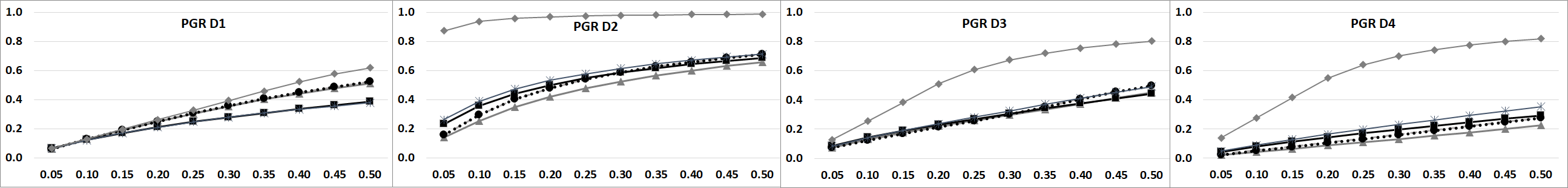}
\vspace{-8pt}
\caption{Performance of Composite Dynamic Progressive GIA.nt (\textsf{CDPG}), Progressive GIA.nt with $JS$ (\textsf{PGJS}) and $MBRO$ (\textsf{PGMB}), Dynamic Progressive GIA.nt with $JS$ (\textsf{DPGJS}) and $MBRO$ (\textsf{DPGMB}) as well as the optimal approach (\textsf{OPTI}) over $D_1$-$D_4$ using as budgets all portions of candidate pairs in $[0.05, 0.50]$ with a step of $0.05$.}
\vspace{-12pt}
\label{fig:progressiveExp}
\end{figure*}

\textit{Overall, Geospark, Location Spark and Parallel GIA.nt are the most time efficient batch algorithms. The first two depend heavily on their parameter configuration, unlike Parallel GIA.nt, whose performance is quite robust with its default configuration,} ranking at least as the second best algorithm in all datasets, but $D_6$.

\vspace{3pt}
\noindent
\textbf{Serial budget-aware processing.} To assess the relative performance of the main progressive algorithms, we compare the best ones over $D_1$-$D_4$. These algorithms are 
progressive algorithm from \cite{DBLP:conf/www/0001MMK21}: 
Progressive GIA.nt in combination with the Jaccard and the $MBRO$ similarity weighting schemes, denoted by \textsf{PGJS} and \textsf{PGMB}, respectively, their dynamic counterparts, \textsf{DPGJS} and \textsf{DPGMB} as well as Composite Dynamic Progressive GIA.nt, \textsf{CDPG}, which uses $JS$ as the primary scheme and $MBRO$ as the secondary one to break the ties. As baseline, we use the optimal progressive algorithm (\textsf{OPTI}), which verifies all topologically related pairs before the non-related ones. For every dataset, we report the performance with respect to all effectiveness for all budgets in the interval $[0.05 \cdot |C|, 0.50 \cdot |C|]$ with a step of 0.05, where $|C|$ denotes the set of candidate pairs. Looking into the results, which appear in Figure \ref{fig:progressiveExp}, we observe the following patterns:

$\bullet$ In $D_1$, the methods involving the $JS$ weighting scheme, i.e., \textsf{CDPG}, \textsf{PGJS} and \textsf{DPGJS}, outperform those relying on $MBRO$, \textsf{PGMB} and \textsf{DPGMB}. This should be attributed to the large proportion of the related pairs that satisfy the relation \texttt{Touches} ($\sim$64.6\% in Table \ref{tb:spatialDatasets}), which is hard to be detected by the latter weighting scheme; the bounding rectangles of touching geometries typically have very low overlap and, thus, $MBRO$ yields extremely low weights for the corresponding related pairs. Nevertheless, all methods achieve very high performance that is close to the optimal one, due to the relatively large portion of qualifying pairs: $\sim$38\% of all candidate pairs are topologically related, as shown in Table \ref{tb:spatialDatasets}.

In more detail, the highest precision and recall is achieved by \textsf{PGJS} and \textsf{DPGJS}, with \textsf{CDPG} following in very close distance. Their distance from the optimal approach increases in proportion to the budget, starting from 0.7\% for $BU$=0.05$\cdot|C|$ and raising up to 16.9\% for $BU$=0.50$\cdot|C|$. On average, across all budgets, the optimal approach outperforms these methods by 13.3\% with respect to both measures. Regarding PGR, the best performance is consistently achieved by \textsf{CDPG}, with \textsf{PGJS} and \textsf{DPGJS} following in close distance. Its distance from the optimal performance increases with the budget, but to a lesser extent than precision and recall. On average, it amounts to just 8.3\%.

$\bullet$ In $D_2$, the relative performance of the considered methods depends on the budget. For precision and recall, the two methods leveraging $MBRO$, \textsf{PGMB} and \textsf{DPGMB}, outperform all others for small budgets up to $BU$=0.25$\cdot|C|$. For $BU$=0.30$\cdot|C|$, all methods achieve practically identical performance. For larger budgets, \textsf{PGJS} and \textsf{DPGJS} achieve the top performance, outperforming \textsf{CDPG} only by an insignificant extent. On average, the distance of the best progressive method from the optimal one amounts to 26.7\% for both measures. This is double than the corresponding distances in $D_1$, a situation that should be attributed to the heavy class imbalance in $D_2$: just 1.3\% of all candidate pairs are qualifying.

Regarding PGR, \textsf{DPGMB} outperforms all other algorithms to a significant extent. \textsf{PGMB} lies in the second place, with its PGR being lower by 5.8\%, on average, across all budgets. \textsf{DPGJS} and \textsf{CDPG} exhibit similar behaviors, underperforming \textsf{DPGMB} by 10.7\%, on average. \textsf{PGJS} yields the worst performance, which is lower by 20\% than \textsf{DPGMB}, on average. Note that in all cases, the smaller the budget is, the higher are differences with \textsf{CDPG}. Note also that the distance of \textsf{DPGMB} from the optimal approach is very high, i.e, $\sim$42.5\% on average, due to the heavy class imbalance.

$\bullet$ In $D_3$, we observe similar patterns as in $D_2$. The recall and precision of \textsf{PGMB} and \textsf{DPGMB} is much higher than the rest of the methods for budgets up to $BU$=0.25$\cdot|C|$, while their PGR is consistently higher across all budgets. The average distance of the best progressive approach from the optimal one is significantly higher than $D_1$ and $D_2$, exceeding 40\% for all evaluation measures. This should be attributed to the topological relations that dominate the qualifying pairs in $D_3$, but are underrepresented or absent from $D_1$ and $D_2$: \texttt{CoveredBy}, \texttt{Overlaps}, \texttt{Within} and \texttt{Equals}.

$\bullet$ In $D_4$, \textsf{PGMB} and \textsf{DPGMB} lie between the optimal approach and the progressive methods that leverage $JS$. In fact, their average distance from the former exceeds 64\% for all evaluation measures, while they outperform \textsf{CDPG}, \textsf{DPGJS} and \textsf{PGJS} by 27.7\% and 33.8\% in terms of precision/recall and PGR, respectively. The high distance from the ideal solution is caused by the same types of topological relations as in $D_3$. The superiority of $MBRO$ over $JS$ is caused by the polygons that are exclusively contained in $D_4$.

It is worth noting at this point that the strength of progressive methods is demonstrated in the evolution of recall in Figure \ref{fig:progressiveExp} over $D_1$, $D_2$ and $D_3$. For $BU$=0.50$\cdot|C|$, the top performing approach has detected 83.1\%, 90.4\% and 82.2\% of all qualifying pairs, respectively. The only exception is $D_4$, where the number of detected qualifying pairs increases linearly with the size of the budget, as its recall amounts to 53.0\% for $BU$=0.50$\cdot|C|$.
\section{Qualitative Analysis}
\label{sec:qualitativeAnalysis}

\begin{figure}[t]
\centering
\includegraphics[width=0.4\textwidth]{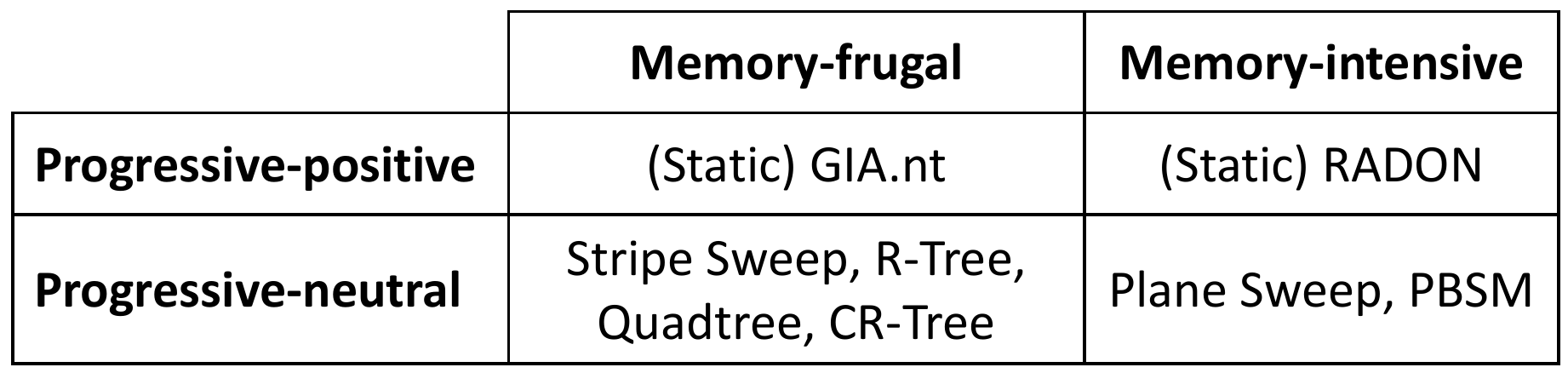}
\vspace{-8pt}
\caption{Taxonomy of serial, budget-agnostic algorithms.}
\vspace{-20pt}
\label{fig:bTaxonomy}
\end{figure}

In this section, we discuss the time and space complexity of JedAI-spatial's budget-agnostic algorithms to provide more insights into the experimental results of Section~\ref{sec:quantAnalysis}.

Figure \ref{fig:bTaxonomy} defines a novel, two-dimensional taxonomy, where the horizontal axis distinguishes the algorithms into \textit{memory-frugal} and \textit{memory-intensive} ones. The former read the target geometries from the disk on the fly, whereas the latter load them into main memory a-priori. This means that the memory-frugal algorithms maintain only the source geometries in main memory, having a space complexity of $O(|S|)$, unlike the memory-intensive ones, whose memory requirements amount to $O(|S|+|T|)$. Hence, the former scale to much larger datasets than the latter. 

The vertical axis of our taxonomy categorizes the algorithms into
\textit{progressive-positive} and \textit{progressive-neutral} ones, depending on whether they provide evidence for budget-aware methods. This is true only for the grid-based methods, which allow for estimating the number of tiles that intersect both geometries in a pair -- this information lies at the core of most weighting schemes in Section \ref{sec:progressiveSerial}. These weighting schemes provide quick and generic evidence about the likelihood that two geometries satisfy at least one topological relation, as verified in \cite{DBLP:conf/www/0001MMK21}. The rest of the budget-agnostic algorithms are not compatible with such weighting schemes and with a progressive functionality, in general.

Another important aspect of budget-agnostic algorithms is the time complexity of their Filtering and Verification. The former is dominated by the cost of constructing the index, 
while the latter is dominated by the cost of processing the candidate pairs. These complexities are summarized in Table \ref{tb:complexityTable}, assuming the average case, where the partitioning of data into grid cells or stripes is not skewed, R-Tree is properly configured and the majority of the non-leaf nodes in Quadtree have four children. Note that $C$ stands for the set of candidate pairs, i.e., source and target geometries with intersecting MBRs: $C =\{(s,t) : MBR(s) \cap MBR(t) \neq \{\}|\}$.

We observe that the most efficient Filtering corresponds to Static GIA.nt, GIA.nt and Stripe Sweep. The first approach iterates once over the source geometries to index them according to the user-defined Equigrid, whereas the other two algorithms perform two iterations: one for extracting the grid granularity from the average dimensions of the source geometries and one for indexing them. The cost of indexing an individual geometry $g$ is considered constant, $O(1)$, for all three algorithms, as the tiles intersecting $MBR(g)$ are determined through simple computations.

The next most efficient Filtering corresponds to the tree-based algorithms. They index only the source geometries, inserting them into the tree one by one. For R-Tree, the cost of a single insertion is $O(\log_M|S|)$, where $M$ denotes the average node capacity. For Quadtree, it amounts to $O(\log_4|S|)$, assuming that every node has exactly four children. For CR-Tree, JedAI-spatial uses Sort Tile Recursive (STR) packing \cite{DBLP:conf/icde/LeuteneggerEL97} to minimize the high space utilization of R-Tree. This approach sorts all source geometries by $x_{min}$, partitions them into vertical slices, sorts them by $y_{min}$ and packs them into tree nodes. Overall, its time complexity is $O(|S| \log |S|)$.

These algorithms are followed by Static RADON and RADON. The former iterates once over all source and target geometries to index them using a predefined Equigrid, while the latter involves one more iteration, which extracts the grid dimensions from the average width and height of the geometries' MBR.

Finally, the least efficient Filtering corresponds to Plane Sweep, which sorts both the source and the target geometries in increasing $x_{min}$. The same process is applied by PBSM inside every one of the user-defined partitions, yielding the same overall time complexity. In our implementation, we use Quicksort \cite{DBLP:journals/cj/Hoare62} to minimize the cost of sorting for both algorithms.

Regarding Verification, its time complexity is determined by two factors: (i) the cost of computing the Intersection Matrix for all candidate pairs with intersecting MBRs, which is denoted by $O(|C|)$ and is common to all algorithms, and (ii) the cost of retrieving these pairs from the index constructed during Filtering. The latter applies only to memory-frugal algorithms, where the target geometries are available only during Verification, unlike the memory-intensive ones. In more detail, 
(Static) GIA.nt and Stripe Sweep generate the candidate pairs with a linear cost, $O(|T|)$, which is required for reading the target geometries from the disk -- the cost of computing the tiles or stripes intersecting the MBR of an individual target geometry is constant, $O(1)$, involving simple computations.
In practice, the reading overhead is rather low, especially when using an SSD, instead of a mechanical hard disk. The tree-based algorithms involve the same overhead, increasing it by the cost of querying the index for retrieving the candidate source geometries. For each target geometry, this cost is $O(\log_M|S|)$ for R-Tree and CR-Tree ($M$ is the average node capacity) and $O(\log_4|S|)$ for Quadtree.

The above theoretical analysis verifies the patterns observed in Figures \ref{fig:filteringScalability} and \ref{fig:verificationScalability}, as our experiments satisfy the underlying assumption (i.e., relatively uniform partitioning of data into tiles and stripes, fine-tuned R-Tree and proper structure for Quadtree).

\begin{table}[t]\centering
\small
    \caption{Time complexity for Filtering and Verification of JedAI-spatial's serial, budget-agnostic algorithms.}
    \vspace{-10pt}
	\begin{tabular}{ | l | c | c | }
        \cline{2-3}
        \multicolumn{1}{c|}{} & Filtering & Verification \\
		\hline
		\hline
		(Static) RADON & $O(|S|$+$|T|)$ & $O(|C|)$ \\
		(Static) GIA.nt & $O(|S|)$ & $O(|C|$+$|T|)$ \\
		\hline
		Plane Sweep & $O(|S| \log |S| $+$ |T| \log |T| )$ & $O(|C|)$\\
		PBSM & $O(|S| \log |S| $+$ |T| \log |T| )$ & $O(|C|)$\\
		Stripe Sweep & $O(|S|)$ & $O(|C|$+$|T|)$ \\
		\hline
		R-Tree & $O(|S| \log_{M} |S|)$ & $O(|C| $+$ |T| \log_{M} |S|)$\\
		Quadtree & $O(|S| \log_4 |S|)$ & $O(|C| $+$ |T| \log_4 |S|)$\\
		CR-Tree & $O(|S| \log |S|)$ & $O(|C| $+$ |T| \log_{M} |S|)$ \\
		\hline
	\end{tabular}
	\vspace{-18pt}
	\label{tb:complexityTable}
\end{table}
\section{Conclusions}
We presented JedAI-spatial, an open-source system that acts as a library of the state-of-the-art algorithms for Geospatial Interlinking. For the existing algorithms, some of which have not been applied to Geospatial Interlinking before, JedAI-spatial incorporates optimized implementations. It also includes new, high-performing techniques. JedAI-spatial facilitates the application of these methods by offering two wizard-like interfaces that assume no expert knowledge and by organizing them into a novel 3D taxonomy that allows users to a-priori select the dimensions of interest (and the corresponding techniques), based entirely on the application requirements. Their benchmarking functionality allows for evaluating the relative performance of the available algorithms and for examining the impact of their internal configuration on their performance. We elaborated on JedAI-spatial's architecture, describing the components of its back- and front-end, and performed thorough experimental analyses, highlighting the relative performance of all serial and parallel budget-agnostic algorithms. 

In the future, we plan to extend JedAI-spatial with support for proximity relations and~point~geometries. 

\bibliographystyle{abbrv}
\bibliography{references}

\end{document}